\documentclass[aps,prb,reprint,groupedaddress]{revtex4-2}

\usepackage{amsmath,amssymb}
\usepackage{graphicx}
\usepackage{xcolor}
\usepackage{physics}
\usepackage[bookmarksnumbered]{hyperref}

\begin{document}

\title{Proof of absence of local conserved quantities in the mixed-field Ising chain}

\author{Yuuya Chiba}
\email{chiba@as.c.u-tokyo.ac.jp}
\affiliation{Department of Basic Science, The University of Tokyo, 3-8-1 Komaba, Meguro, Tokyo 153-8902, Japan}

\date{\today}

\begin{abstract}
Absence of local conserved quantities is often required, such as for thermalization or for the validity of response theory.  Although many studies have discussed whether thermalization occurs in the Ising chain with longitudinal and transverse fields, rigorous results on local conserved quantities of this model have still been lacking.  Here, we rigorously prove that, if all coupling constants are nonzero, this model has no conserved quantity spanned by local operators with support size up to half of the system size other than a trivial one, i.e., a linear combination of the Hamiltonian and the identity.  The proof is given not only for the periodic boundary condition but also for the open boundary condition.  We also discuss relation to the integrability of the model where the longitudinal field is set to zero.  Our results provide the second example of spin models whose nonintegrability is rigorously proved.  

\end{abstract}

\maketitle

\section{\label{sec:Introduction}Introduction}

In a quantum many-body system,
the local conserved quantities
impose
significant constraints
on the unitary dynamics.
For instance,
in the context of thermalization in isolated quantum systems~\cite{Neumann1929,Neumann2010,Deutsch1991,Srednicki1994,Rigol2008,DAlessio2016,Mori2018},
it is known that we need to specify all local conserved quantities
to describe the relaxed state correctly~\cite{Rigol2008,Mori2018}.
On the other hand,
thermodynamics assumes that
the equilibrium state is specified by
a set of variables whose number is independent of the system size.
Thus,
for the unitary dynamics to be consistent with thermodynamics,
we need to confirm
that the number of local conserved quantities
is independent of the system size.

In addition, 
chaos in quantum many-body systems
has attracted much attention
due to its relation 
with thermalization~\cite{Neumann1929,Neumann2010,Deutsch1991,Srednicki1994,Rigol2008}
and quantum information theory~\cite{Hayden2007,Sekino2008,Shenker2014,Hosur2016,Yoshida2019,Zhu2022}.
When testing whether a quantum many-body system is
chaotic,
one often examines
energy-level spacing distribution~\cite{
Wigner1955,Dyson1962,Dyson1962a,Mehta2004,
Oganesyan2007,Atas2013,Giraud2022,
Santos2010,Kollath2010,Mondaini2016
}.
However,
we need to construct this distribution
specifying
all local conserved quantities and symmetries;
otherwise,
the observed distribution will never be described by
the Wigner-Dyson distribution~\cite{Giraud2022},
which is the evidence of chaos.
Hence,
finding all local conserved quantities
is important 
for checking the signature of quantum chaos.

Furthermore,
the validity of linear response theory~\cite{Green1952,Kubo1957,KTH}
has long been discussed~\cite{Kubo1957,Mazur1969,Suzuki1971,KTH,Shimizu1999,Endo2018,Chiba2020,Chiba2023}.
It has been shown that,
for the response
to be consistent with thermodynamics
in the static limit,
absence of local conserved quantities (that have finite overlaps with the observable of interest) is necessary~\cite{Mazur1969,Suzuki1971}.

By contrast,
it has been considered that integrable systems have an extensive number of local conserved quantities.
For instance,
in integrable systems solvable by the Bethe ansatz~\cite{Bethe1931,Yang1966,Sklyanin1979,Takhtadzhan1979,Baxter1982,Korepin1993,Takahashi1999},
an extensive number of local conserved quantities have been constructed from the transfer matrix~\cite{Baxter1982,Korepin1993,Takahashi1999}.
As a result,
they do not thermalize~\cite{Rigol2007,Wouters2014,Pozsgay2014,Ilievski2015a,Doyon2017}
and do not exhibit quantum chaotic behaviors~\cite{Santos2010,Kollath2010}.
Furthermore,
Zotos \textit{et al.}~\cite{Zotos1997,Zotos1999} observed
an anomalous behavior of the response of the spin current
in an integrable spin system~\cite{Zotos1997,Zotos1999},
indicating
the existence of additional conserved quantities
called quasilocal conserved quantities~\cite{Ilievski2015,Mierzejewski2015,Ilievski2016}.
This discovery has stimulated further research on 
conserved quantities in integrable systems~\cite{Fagotti2016,Piroli2017,Zadnik2017,DeLuca2017,Palmai2018,Ljubotina2019,Nozawa2020}.

Despite the importance explained above,
conserved quantities in nonintegrable systems
have been analyzed only numerically or approximately.
For instance,
in the Ising chain with longitudinal and transverse fields (the mixed-field Ising chain),
the level spacing distribution~\cite{Banuls2011}
and the condition for thermalization 
called
eigenstate thermalization hypothesis~\cite{Kim2014}
have been studied only numerically.
As a result,
these studies just suggest
but do not prove absence of local conserved quantities.

Furthermore,
Ba\~{n}uls \textit{et al.} found that
the above-mentioned model relaxed to a nonthermal state and remained in that state
for some initial states 
until the timescale they could simulate~\cite{Banuls2011}.
In addition,
certain observables and the entanglement entropy
oscillate in time
for a very long time in this model~\cite{Banuls2011,Kormos2017,Castro-Alvaredo2020},
indicating absence of thermalization.
These results seem to contradict
the above numerical results~\cite{Kim2014} that suggest
existence of thermalization.
This paradox has been tackled 
in Ref.~\cite{Kim2015}
by numerically constructing operators that are approximately conserved.
However,
since all above results are mainly based on numerical calculations,
they suffer from finite-size effects or numerical errors.
Thus, rigorous results on conserved quantities of this model
has been desired~\cite{Banuls2011,Kim2014,Kim2015,Kormos2017,Castro-Alvaredo2020,Xu2020,Wurtz2020}.

Exceptionally, 
Shiraishi rigorously proved that
the XYZ spin chain with a magnetic field
does not have local conserved quantities~\cite{Grabowski1995,Shiraishi2019}.
However, his result is not applicable to 
the above-mentioned model,
and extension to 
that model
is not 
obvious.

In this paper,
we investigate local conserved quantities 
in the Ising chain with longitudinal and transverse fields,
and prove their absence rigorously.
We show that,
if all coupling constants in the Hamiltonian are nonzero,
any conserved quantity
spanned by
local operators with support size up to half the system size $N/2$
is a trivial one:
a linear combination of the Hamiltonian and the identity.
This result holds both for the periodic boundary condition and the open boundary condition (OBC).

First, we prove the above result in the case of the periodic boundary condition.
Thanks to the translation invariance,
local conserved quantities can be taken as eigenoperators of translation,
without loss of generality.
Suppose that the local conserved quantity under consideration
is spanned by the products 
of matrices $\{I,X,Y,Z\}$ (the identity and the Pauli matrices)
on consecutive sites up to $L$ sites.
Our proof can be divided into four parts.
The first to third parts analyze a local conserved quantity with $3\le L\le N/2$.
The first part shows that 
its expansion coefficients in the basis corresponding to the products of $L$ matrices where neither end is $Z$ are zero.
The second part shows that 
those
where one end is $Z$
while the other is not $Z$ are zero.
The third part shows that 
those
where both ends are $Z$ are zero.
Combining these three parts,
we show that there is no local conserved quantity with $3\le L\le N/2$.
The fourth part analyzes a local conserved quantity with $L\le 2$
and shows that it is restricted to a linear combination of the Hamiltonian and the identity.

Next we prove the above result in the case of the OBC.
Because the OBC breaks the translation invariance,
we need to extend the above proof.
The key to the extension is the fact that
the proof remains almost unchanged,
even if we replace the eigenoperators of translation in the proof for the periodic boundary condition,
with site-dependent operators, which are not eigenoperators of translation. 

After that,
we explain
why our proof is not applicable 
when the support size $L$ is larger than $N/2$,
and discuss
the possibility of extending our main result to larger support sizes.
We also discuss
the relation between our results
and 
the existence of local conserved quantities 
in the transverse-field Ising chain 
(the model where longitudinal field is taken to zero).

The paper is organized as follows.
Section~\ref{sec:Setup} explains our setup
and introduces the definition of local conserved quantities.
Our main result,
i.e., the absence of local conserved quantities, 
is described in Sec.~\ref{sec:Result} 
and proved in Sec.~\ref{sec:Proof}.
These Secs.~\ref{sec:Setup}--\ref{sec:Proof} deal with the periodic boundary condition. Extension to the OBC is given in Sec.~\ref{sec:OBC}.
Section~\ref{sec:Discussion} discusses 
the possibility of extension to larger support sizes
and relation to the integrability of the transverse-field Ising chain.
Section~\ref{sec:Summary} gives the summary of the paper.

\section{\label{sec:Setup}Setup}

We consider
the Ising model with longitudinal and transverse fields
on a one-dimensional lattice with $N$ sites,
\begin{align}
\hat{H}=-\sum_{j=1}^{N}\bigl(
J\hat{Z}_{j}\hat{Z}_{j+1}
+h_{z}\hat{Z}_{j}
+h_{x}\hat{X}_{j}
\bigr),
\label{eq:Hamiltonian}
\end{align}
where $\hat{A}_{j}$ 
for $A=X,Y,Z$ is the Pauli operator
acting on the site $j$.
Here we impose the periodic boundary condition 
$\hat{A}_{j+N}=\hat{A}_{j}$.
(The case of the OBC is analyzed in Sec.~\ref{sec:OBC}.)

We explain the definition of a local conserved quantity.
First,
we introduce the \emph{$\ell$-support basis} by
\begin{align}
\hat{\vb*{A}^{\ell}}_{j}
=\hat{A^{1}}_{j}\hat{A^{2}}_{j+1}...\hat{A^{\ell}}_{j+\ell-1},
\label{eq:basis_A^ell}
\end{align}
with 
\begin{align}
A^{1},A^{\ell}&\in\{X,Y,Z\},
\notag\\
A^{2},...,A^{\ell-1}&\in\{I,X,Y,Z\},
\label{eq:Range_A}
\end{align}
and the symbol $I$ represents the identity.
Here we impose the periodic boundary condition.
In the following, unless otherwise stated,
$\vb*{A}^{\ell}$ represents a sequence $A^{1}A^{2}...A^{\ell}$
satisfying Eq.~(\ref{eq:Range_A}).
For a given $j$ and $\ell$, the set $\{\hat{\vb*{A}^{\ell}}_{j}\}_{\vb*{A}^{\ell}}$ is a basis of operators supported on the consecutive sites $j,j+1,...,j+\ell-1$.
Using these, we say that an operator $\hat{Q}$ is an $L$-local conserved quantity
when it is spanned by the $\ell$-support basis for all $j$ and $\ell\le L$,
\begin{align}
\hat{Q}
=\sum_{\ell=1}^{L}
\sum_{j=1}^{N}
\sum_{\vb*{A}^{\ell}}
c_{\vb*{A}^{\ell}_{j}}
\hat{\vb*{A}^{\ell}}_{j}
+c_{I}
\hat{I},
\label{eq:L-local}
\end{align}
and commutes with $\hat{H}$.
(Here $c_{\vb*{A}^{\ell}_{j}},c_{I}\in\mathbb{C}$ are the expansion coefficients.)

Let $\hat{\mathcal{T}}$ be the translation operator 
that translates states by one site.
Because of the translational symmetry of the Hamiltonian Eq.~(\ref{eq:Hamiltonian}),
any $L$-local conserved quantity $\hat{Q}$ can be
written as a linear combination of
$L$-local conserved quantities $\{\hat{Q}^{(q)}\}_{q}$
that 
satisfy
\begin{align}
\hat{\mathcal{T}}\hat{Q}^{(q)}\hat{\mathcal{T}}^{\dagger}
=e^{iq}\hat{Q}^{(q)},
\end{align}
with $q=2\pi n/N$ and $n=0,1,...,N-1$.
This means that
$\hat{Q}^{(q)}$
is an eigenoperator of $\hat{\mathcal{T}}$
with an eigenvalue $e^{iq}$.

Now we introduce a basis of eigenoperators of $\hat{\mathcal{T}}$
with an eigenvalue $e^{iq}$ as
\begin{align}
\hat{M}^{(q)}_{\vb*{A}^{\ell}}
=\sum_{j=1}^{N}e^{-iqj}
\hat{\vb*{A}^{\ell}}_{j}.
\label{eq:M^q}
\end{align}
Then we can expand
such conserved quantities $\hat{Q}^{(q)}$
in this basis 
as
\begin{align}
\hat{Q}^{(q)}
=\sum_{\ell=1}^{L}
\sum_{\vb*{A}^{\ell}}
c^{(q,\ell)}_{\vb*{A}^{\ell}}
\hat{M}^{(q)}_{\vb*{A}^{\ell}}
+c^{(q=0,0)}_{I}\delta_{q,0}
\hat{I}
\label{eq:Q^q}
\end{align}
with expansion coefficients $\{c^{(q,\ell)}_{\vb*{A}^{\ell}}\}_{\vb*{A}^{\ell}}$.
(Here we added $\ell$ to the superscript to emphasize its value.)
Thus 
we explore
whether the model Eq.~(\ref{eq:Hamiltonian})
has a nontrivial solution $\{c^{(q,\ell)}_{\vb*{A}^{\ell}}\}_{\vb*{A}^{\ell}}$ 
to
\begin{align}
[\hat{Q}^{(q)},\hat{H}]=0.
\label{eq:[Q,H]=0}
\end{align}

\section{\label{sec:Result}Main Result}

We show that
for $J,h_z,h_x\neq 0$,
and $L\le N/2$~\footnote{
The reasons for the restriction $L\le N/2$ are explained
in Sec.~\ref{sec:Discussion_LargerL}.
},
there is no $L$-local conserved quantity
that is linearly independent of $\hat{H}$ and $\hat{I}$.
In other words,
the solution of Eq.~(\ref{eq:[Q,H]=0}) 
satisfies
\begin{align}
&c^{(q,\ell)}_{\vb*{A}^{\ell}}=0 \qquad 
\text{for any }c^{(q,\ell)}_{\vb*{A}^{\ell}}
\notag\\
&\text{ other than }
c^{(q=0,0)}_{I}, c^{(q=0,2)}_{ZZ}, c^{(q=0,1)}_{Z}, c^{(q=0,1)}_{X}
\label{eq:Result1}
\end{align}
and
\begin{align}
c^{(q=0,2)}_{ZZ}/J
=c^{(q=0,1)}_{Z}/h_z
=c^{(q=0,1)}_{X}/h_x.
\label{eq:Result2}
\end{align}
This means that
any $L$-local conserved quantity $\hat{Q}^{(q)}$
can be written as
\begin{align}
\hat{Q}^{(q)}=\delta_{q,0}\bigl(c^{(q=0,0)}_{I}\hat{I}-c^{(q=0,2)}_{ZZ}\hat{H}/J\bigr).
\label{eq:Result3}
\end{align}

In the model Eq.~(\ref{eq:Hamiltonian}),
behaviors suggesting the absence of local conserved quantities
have been previously observed.
For example,
the level spacing distribution
has been described by the Wigner-Dyson distribution~\cite{Banuls2011},
and the eigenstate thermalization hypothesis has been verified 
for some observables~\cite{Kim2014}.
Our result proves this suggestion rigorously,
and is completely consistent with these behaviors~\footnote{In studies of quantum many-body systems, a nonintegrable system usually refers to a system that has only a finite number of local conserved quantities~\cite{Mori2018}.
Our main result shows a stronger statement that local conserved quantities of the model are restricted to only trivial ones.
This seems to be related to the famous argument of Parke~\cite{Parke1980} in $1+1$ dimensional quantum field theory, which shows that existence of two nontrivial conserved charges implies solvability of the S matrix.}.

Note that, 
since we analyze all values of $q$,
the above result also shows that
any $L$-local conserved quantity of the form Eq.~(\ref{eq:L-local})
can be written as a linear combination of $\hat{H}$ and $\hat{I}$
for $L\le N/2$.

Note also that,
in the definition of the Hamiltonian $\hat{H}$
and the $L$-local conserved quantity $\hat{Q}^{(q)}$,
the periodic boundary condition is imposed.
We extend these to the OBC
in Sec.~\ref{sec:OBC}.

\section{\label{sec:Proof}Proof}

In this section,
we prove our main result given in Sec.~\ref{sec:Result}.
In the following, we assume that $L\le N/2$.
Our proof strongly relies on the assumptions $J\neq 0$ and $h_{x}\neq 0$,
but use of $h_{z}\neq 0$ is avoided as much as possible,
so many results can be applied to the case where $h_{z}=0$.
(Such a case is analyzed in Sec.~\ref{sec:Discussion_TFIM} in detail.)

For a candidate of 
the $L$-local conserved quantity $\hat{Q}^{(q)}$,
we define coefficients $\{r^{(q,\ell)}_{\vb*{A}^{\ell}}\}_{\vb*{A}^{\ell}}$
as
\begin{align}
\frac{1}{2i}[\hat{Q}^{(q)},-\hat{H}]
=\sum_{\ell}
\sum_{\vb*{A}^{\ell}}
r^{(q,\ell)}_{\vb*{A}^{\ell}}
\hat{M}^{(q)}_{\vb*{A}^{\ell}}.
\label{eq:r^q}
\end{align}
Using these coefficients,
Eq.~(\ref{eq:[Q,H]=0}) is equivalent to
\begin{align}
r^{(q,\ell)}_{\vb*{A}^{\ell}}=0
\qquad\text{for all }\vb*{A}^{\ell}.
\label{eq:r^q=0}
\end{align}

Our proof is divided into four parts,
which correspond to Secs.~\ref{sec:Proof_noZ}--\ref{sec:Proof_L<=2}.
(See also Table~\ref{tbl:ProofOutline}.)
These parts analyze the coefficients $\{c^{(q,\ell)}_{\vb*{A}^{\ell}}\}_{\ell,\vb*{A}^{\ell}}$
satisfying Eq.~(\ref{eq:r^q=0})
and show that the coefficients must be zero except for a few trivial ones.
The first to the third parts examine
the coefficients with the largest locality $\ell=L$
in the case of $3\le L\le N/2$.
The first part (Sec.~\ref{sec:Proof_noZ}) shows that
the coefficients $c^{(q,L)}_{\vb*{A}^{L}}$
whose $A^{1}$ and $A^{L}$ are non-$Z$ are zero.
The second part (Sec.~\ref{sec:Proof_oneZ}) shows that
the coefficients $c^{(q,L)}_{\vb*{A}^{L}}$
in which one of $A^{1}$ and $A^{L}$ is $Z$ and the other is non-$Z$ are zero.
The third part (Sec.~\ref{sec:Proof_bothZ}) shows that
the coefficients $c^{(q,L)}_{\vb*{A}^{L}}$
whose $A^{1}$ and $A^{L}$ are both $Z$ are zero.
From these three parts, we can say that
there is no $L$-local conserved quantity for $3\le L\le N/2$.
Finally, 
the fourth part (Sec.~\ref{sec:Proof_L<=2}) shows that,
in the case of $L\le 2$, 
the coefficients $c^{(q,2)}_{\vb*{A}^{2}}$ and $c^{(q,1)}_{\vb*{A}^{1}}$ are zero
except for trivial ones of the forms~(\ref{eq:Result1}) and (\ref{eq:Result2}).

\begin{table}
\caption{\label{tbl:ProofOutline}%
Structure of the proof.} 
\begin{ruledtabular}
\begin{tabular}{lcc}
Section & What is proved & Range of $L$
\\ \hline
\ref{sec:Proof_noZ} & $c^{(q,L)}_{X...X}=c^{(q,L)}_{X...Y}=c^{(q,L)}_{Y...X}=c^{(q,L)}_{Y...Y}=0$ & $2\le L\le N/2$
\\ 
\ref{sec:Proof_oneZ} & $c^{(q,L)}_{Z...X}=c^{(q,L)}_{Z...Y}=c^{(q,L)}_{X...Z}=c^{(q,L)}_{Y...Z}=0$ & $3\le L\le N/2$
\\ 
\ref{sec:Proof_bothZ} & $c^{(q,L)}_{Z...Z}=0$ & $3\le L\le N/2$
\\ 
\ref{sec:Proof_L<=2} & $c^{(q,2)}_{...},\ c^{(q,1)}_{...}=\text{trivial}$ & $L \le 2$
\end{tabular}
\end{ruledtabular}
\end{table}%

\subsection{\label{sec:Proof_noZ}Coefficients \texorpdfstring{$c^{(q,L)}_{\vb*{A}^{L}}$}{c L} where neither end is $Z$}

To represent
$\{r^{(q,\ell)}_{\vb*{A}^{\ell}}\}_{\vb*{A}^{\ell}}$ 
in terms of 
$\{c^{(q,\ell)}_{\vb*{A}^{\ell}}\}_{\vb*{A}^{\ell}}$,
we introduce a graphical notation of Ref.~\cite{Shiraishi2019}
as follows.
For instance,
if $L\ge 3$, $\hat{Q}^{(q)}$ represented by Eq.~(\ref{eq:Q^q}) includes
the term $c^{(q,3)}_{XZY}\hat{M}^{(q)}_{XZY}$,
and hence the left-hand side (LHS) of Eq.~(\ref{eq:r^q})
includes
the following term
\begin{align}
&\frac{1}{2i}[c^{(q,3)}_{XZY}\hat{M}^{(q)}_{XZY},J\hat{M}^{(q=0)}_{ZZ}]\notag\\
&=J c^{(q,3)}_{XZY} \sum_{j=1}^{N}\frac{e^{-iqj}}{2i}
\notag\\
&\quad\times\bigl(
[\hat{X}_{j}\hat{Z}_{j+1}\hat{Y}_{j+2},\hat{Z}_{j-1}\hat{Z}_{j}]
+[\hat{X}_{j}\hat{Z}_{j+1}\hat{Y}_{j+2},\hat{Z}_{j}\hat{Z}_{j+1}]
\notag\\
&\quad
+[\hat{X}_{j}\hat{Z}_{j+1}\hat{Y}_{j+2},\hat{Z}_{j+1}\hat{Z}_{j+2}]
+[\hat{X}_{j}\hat{Z}_{j+1}\hat{Y}_{j+2},\hat{Z}_{j+2}\hat{Z}_{j+3}]
\bigr)
\label{eq:Graphical_ex1}
\\
&=J c^{(q,3)}_{XZY}
\bigl(-e^{-iq}\hat{M}^{(q)}_{ZYZY}-\hat{M}^{(q)}_{YIY}
+\hat{M}^{(q)}_{XIX}+\hat{M}^{(q)}_{XZXZ}\bigr).
\label{eq:Graphical_ex2}
\end{align}
Thus, these four terms contribute
to $r^{(q,4)}_{ZYZY}$, $r^{(q,3)}_{YIY}$, $r^{(q,3)}_{XIX}$, and $r^{(q,4)}_{XZXZ}$.
We represent
such contributions by the following diagrams:
\begin{align}
\begin{array}{rccc}
&X&Z&Y\\
Z&Z&&\\ \hline
-Z&Y&Z&Y
\end{array}
\quad
\begin{array}{rcc}
 X&Z&Y\\
 Z&Z& \\ \hline
-Y&I&Y
\end{array}
\quad
\begin{array}{rcc}
 X&Z&Y\\
  &Z&Z\\ \hline
 X&I&X
\end{array}
\quad
\begin{array}{rccc}
 X&Z&Y& \\
  & &Z&Z \\ \hline
 X&Z&X&Z
\end{array}.
\label{eq:Graphical_ex3}
\end{align}
In each diagram, 
the first row represents the term from $\hat{Q}^{(q)}$,
the second row the term from $-\hat{H}$,
and the third row the contribution to $[\hat{Q}^{(q)},-\hat{H}]/2i$.

From Eqs.~(\ref{eq:Hamiltonian}) and (\ref{eq:Q^q}),
$[\hat{Q}^{(q)},\hat{H}]$ is spanned by the $\ell$-support basis
for all $j$ and $\ell\le L+1$,
and hence,
\begin{align}
r^{(q,\ell)}_{\vb*{A}^{\ell}}=0
\qquad\text{for }\ell\ge L+2.
\end{align}
Therefore, it is enough to investigate
$r^{(q,\ell)}_{\vb*{A}^{\ell}}$ with $\ell \le L+1$.

We start from $c^{(q,L)}_{\vb*{A}^{L}}$
whose $A^{1}$ and $A^{L}$ are both $X$ or $Y$.
As a first example, 
we consider the following diagram,
which includes the coefficient 
$c^{(q,L)}_{XA^{2}...A^{L-1}Y}$:
\begin{align}
\begin{array}{rccccc}
X&A^{2}&...&A^{L-1}&Y& \\
 &     &   &       &Z&Z\\ \hline
X&A^{2}&...&A^{L-1}&X&Z
\end{array}.
\label{eq:Diagram_L_X_Y}
\end{align}
We can easily see that
this is the only contribution to $r^{(q,L+1)}_{XA^{2}...A^{L-1}XZ}$,
because the contribution to $r^{(q,L+1)}_{\vb*{A}^{L+1}}$ comes 
only from 
the commutator of $\hat{M}^{(q)}_{\vb*{A}^{L}}$ and $\hat{M}^{(q=0)}_{ZZ}$.
Therefore,
we have
\begin{align}
r^{(q,L+1)}_{XA^{2}...A^{L-1}XZ}=J\, c^{(q,L)}_{XA^{2}...A^{L-1}Y}.
\end{align}
If we assume $J\neq 0$,
Eq.~(\ref{eq:r^q=0}) shows that
\begin{align}
c^{(q,L)}_{XA^{2}...A^{L-1}Y}=0.
\end{align}

In a similar manner,
we have\\
\textit{Proposition~A}: 
Assume $J\neq 0$. 
For $2\le L\le N/2$,
the solution of Eq.~(\ref{eq:r^q=0}) satisfies
\begin{align}
c^{(q,L)}_{\vb*{A}^{L}}=0 \quad\text{ for }A^{1}=X,Y\text{ and } A^{L}=X,Y.
\label{eq:PropA}
\end{align}
Here the range of $\vb*{A}^{L}=A^{1}...A^{L}$ is given by Eq.~(\ref{eq:Range_A}).

\subsection{\label{sec:Proof_oneZ}Coefficients \texorpdfstring{$c^{(q,L)}_{\vb*{A}^{L}}$}{c L} where one end is $Z$}

Next we consider the remaining coefficients $c^{(q,L)}_{\vb*{A}^{L}}$.
From Proposition~A, they satisfy $A^{1}=Z$ or $A^{L}=Z$.
For instance,
they appear
in the diagram 
\begin{align}
\begin{array}{rccccc}
Z&A^{2}&...&A^{L-1}&X\\
X&     &   &       & \\ \hline
Y&A^{2}&...&A^{L-1}&X
\end{array}.
\label{eq:Z...X(L)_Y...X(L)}
\end{align}
We emphasize that,
in the resulting coefficient $r^{(q,L)}_{YA^{2}...A^{L-1}X}$,
both ends are not $Z$.
Therefore, $r^{(q,L)}_{YA^{2}...A^{L-1}X}$ 
does not include contributions from $c^{(q,L-1)}_{\vb*{A}^{L-1}}$.
Furthermore,
since the remaining coefficient $c^{(q,L)}_{\vb*{A}^{L}}$
satisfies $A^{1}=Z$ or $A^{L}=Z$,
$r^{(q,L)}_{YA^{2}...A^{L-1}X}$ does not include 
contributions from the remaining $c^{(q,L)}_{\vb*{A}^{L}}$ other than Eq.~(\ref{eq:Z...X(L)_Y...X(L)}).
Hence, we have
\begin{align}
r^{(q,L)}_{YA^{2}...A^{L-1}X}
=h_{x} c^{(q,L)}_{ZA^{2}...A^{L-1}X}=0.
\end{align}
In a similar manner,
we can show that\\
\textit{Lemma~B.1}: 
Assume $J\neq 0$ and $h_{x}\neq 0$. 
For $2\le L\le N/2$,
the solution of Eq.~(\ref{eq:r^q=0}) satisfies
\begin{align}
c^{(q,L)}_{ZA^{2}...A^{L-1}X}=c^{(q,L)}_{XA^{2}...A^{L-1}Z}=0.
\label{eq:LemmaB1}
\end{align}
Here the range of $\vb*{A}^{L}=A^{1}...A^{L}$ is given by Eq.~(\ref{eq:Range_A}).

Thus, the remaining coefficients $c^{(q,L)}_{\vb*{A}^{L}}$
can be classified into three types.
We will examine 
$c^{(q,L)}_{ZA^{2}...A^{L-1}Y}$ and $c^{(q,L)}_{YA^{2}...A^{L-1}Z}$
in detail
in the rest of this subsection,
and $c^{(q,L)}_{ZA^{2}...A^{L-1}Z}$ in the next subsection.
As an example,
we consider
\begin{align}
\begin{array}{rccccc}
Z&A^{2}&...&A^{L-1}&Y& \\
 &     &   &       &Z&Z\\ \hline
Z&A^{2}&...&A^{L-1}&X&Z
\end{array}.
\end{align}
When $A^{2}=I,Y,Z$,
it is the only contribution to
$r^{(q,L+1)}_{ZA^{2}...A^{L-1}XZ}$.
(This is because the contribution of the form Eq.~(\ref{eq:Y...XZ(L)_ZX...XZ(L+1)}) below
is absent for $A^{2}=I,Y,Z$.)
Hence, we have 
\begin{align}
J c^{(q,L)}_{ZA^{2}...A^{L-1}Y}=0\quad \text{for }A^{2}=I,Y,Z.
\end{align}
On the other hand,
when $A^{2}=X$,
there is the other contribution,
\begin{align}
\begin{array}{rcccccc}
 &Y&A^{3}&...&A^{L-1}&X&Z\\
Z&Z&     &   &       & & \\ \hline
Z&X&A^{3}&...&A^{L-1}&X&Z
\end{array},
\label{eq:Y...XZ(L)_ZX...XZ(L+1)}
\end{align}
and hence we have
\begin{align}
0&=r^{(q,L+1)}_{ZXA^{3}...A^{L-1}XZ}\\
&=J(c^{(q,L)}_{ZXA^{3}...A^{L-1}Y}+e^{-iq}c^{(q,L)}_{YA^{3}...A^{L-1}XZ})
\end{align}
In a similar manner,
we obtain the following lemma.\\
\textit{Lemma~B.2}: 
Assume $J\neq 0$ and $h_{x}\neq 0$.
For $3\le L\le N/2$, 
the solution of Eq.~(\ref{eq:r^q=0}) satisfies
\begin{align}
&c^{(q,L)}_{ZA^{2}...A^{L-1}Y}=0\quad \text{for }A^{2}=I,Y,Z
\label{eq:ZI...Y(L)=0}
\\
&c^{(q,L)}_{YA^{2}...A^{L-1}Z}=0\quad \text{for }A^{L-1}=I,Y,Z
\label{eq:Y...IZ(L)=0}
\\
&c^{(q,L)}_{ZXA^{3}...A^{L-1}Y}=-e^{-iq}c^{(q,L)}_{YA^{3}...A^{L-1}XZ}
\label{eq:ZX...Y(L)=Y...XZ(L)}
\end{align}
Here the range of $\vb*{A}^{L}=A^{1}...A^{L}$ is given by Eq.~(\ref{eq:Range_A}).

Now we examine $c^{(q,L)}_{ZXA^{3}...A^{L-1}Y}$
in more detail.
It contributes to $r^{(q,L)}_{YXA^{3}...A^{L-1}Y}$ as
\begin{align}
\begin{array}{rcccccc}
Z&X&A^{3}&...&A^{L-1}&Y\\
X& &     &   &       & \\ \hline
Y&X&A^{3}&...&A^{L-1}&Y
\end{array},
\label{eq:Diagram_ZX...Y(L)_X}
\end{align}
which has the other contribution:
\begin{align}
\begin{array}{rcccccc}
Y&X&A^{3}&...&A^{L-1}&Z\\
 & &     &   &       &X\\ \hline
Y&X&A^{3}&...&A^{L-1}&Y
\end{array}.
\label{eq:Diagram_YX...Z(L)_X}
\end{align}
If we assume $h_{x}\neq 0$, we have
\begin{align}
c^{(q,L)}_{ZXA^{3}...A^{L-1}Y}=-c^{(q,L)}_{YXA^{3}...A^{L-1}Z}.
\label{eq:ZX...Y(L)=YX...Z(L)}
\end{align}
From Eq.~(\ref{eq:Y...IZ(L)=0}),
this quantity becomes zero unless $A^{L-1}=X$.
In the same way, 
using Eqs.~(\ref{eq:ZX...Y(L)=Y...XZ(L)}) and (\ref{eq:ZX...Y(L)=YX...Z(L)}),
we can shift the sequence $A^{3}...A^{L-1}$ 
in the coefficient $c^{(q,L)}_{ZXA^{3}...A^{L-1}Y}$ 
to the left.
Then if there is at least one non-$X$ symbol in $A^{3},...,A^{L-1}$,
Eq.~(\ref{eq:Y...IZ(L)=0}) states that this coefficient is zero.
Combining the phase factor $e^{-iq}$ in Eq.~(\ref{eq:ZX...Y(L)=Y...XZ(L)}),
we have
\begin{align}
&c^{(q,L)}_{ZXA^{3}...A^{L-1}Y}=0
\notag\\
&\quad\text{unless }q=0\text{ and }A^{3}=...=A^{L-1}=X.
\label{eq:ZX...Y(L)=0}
\end{align}
Introducing a shorthand notation
\begin{align}
(A)^k:=\underbrace{AA...A}_{k\text{ times}}\quad \text{for }A=I,X,Y,Z,
\end{align}
we obtain the following lemma.\\
\textit{Lemma~B.3}: 
Assume $J\neq 0$ and $h_{x}\neq 0$.
For $3\le L\le N/2$, 
the solution of Eq.~(\ref{eq:r^q=0}) satisfies
\begin{align}
&c^{(q,L)}_{\vb*{A}^{L}}=0\quad 
&\text{unless }
&(A^{1},A^{L})=(Z,Z), 
\notag\\
& & & q=0\text{ and }\vb*{A}^{L}=Z(X)^{L-2}Y,
\notag\\
& & & q=0\text{ and } \vb*{A}^{L}=Y(X)^{L-2}Z.
\label{eq:LemmaB3}
\end{align}
Here the range of $\vb*{A}^{L}=A^{1}...A^{L}$ is given by Eq.~(\ref{eq:Range_A}).

\subsubsection{Coefficients \texorpdfstring{$c^{(q,L)}_{\vb*{A}^{L}}$}{c L} with \texorpdfstring{$\vb*{A}^{L}=Z(X)^{L-2}Y$, $Y(X)^{L-2}Z$}{A=ZXXXY, YXXXZ}}

To conclude this subsection,
we examine the coefficients $c^{(q=0,L)}_{Z(X)^{L-2}Y}$ and $c^{(q=0,L)}_{Y(X)^{L-2}Z}$.
We consider the following diagram:
\begin{align}
\begin{array}{rcc}
Z&(X)^{L-2}&Y\\
 &         &Z\\ \hline
Z&(X)^{L-2}&X
\end{array},
\label{eq:Diagram_ZXXXY(L)_ZXXXX(L)}
\end{align}
which has the other contribution:
\begin{align}
\begin{array}{rcc}
 &Y&(X)^{L-2}\\
Z&Z&         \\ \hline
Z&X&(X)^{L-2}
\end{array}.
\label{eq:Diagram_YXXX(L-1)_ZXXXX(L)}
\end{align}
Thus, we have
\begin{align}
h_{z} c^{(q=0,L)}_{Z(X)^{L-2}Y}+J c^{(q=0,L-1)}_{Y(X)^{L-2}}=0.
\label{eq:ZXXXY(L)_YXXX(L-1)=0}
\end{align}
The second term of the LHS of this equation
also contributes to
\begin{align}
\begin{array}{rccc}
 Y&(X)^{L-3}&X& \\
  &         &Z&Z\\ \hline
-Y&(X)^{L-3}&Y&Z
\end{array},
\label{eq:Diagram_YXXX(L-1)_ZZ}
\end{align}
which also has the other contributions:
\begin{align}
\begin{array}{rccc}
Z&(X)^{L-3}&Y&Z\\
X&         & & \\ \hline
Y&(X)^{L-3}&Y&Z
\end{array}
\quad
\begin{array}{rccc}
 Y&(X)^{L-3}&X&Z\\
  &         &Z& \\ \hline
-Y&(X)^{L-3}&Y&Z
\end{array}.
\label{eq:Diagram_YXXXZ(L)_YXXYZ(L)}
\end{align}
These result in
\begin{align}
-J c^{(q=0,L-1)}_{Y(X)^{L-2}}
+h_{x} c^{(q=0,L)}_{Z(X)^{L-3}YZ}
-h_{z} c^{(q=0,L)}_{Y(X)^{L-2}Z}
=0.
\end{align}
Combining this with Eqs.~(\ref{eq:ZX...Y(L)=Y...XZ(L)}) and (\ref{eq:ZXXXY(L)_YXXX(L-1)=0}), we have
\begin{align}
h_{x} c^{(q=0,L)}_{Z(X)^{L-3}YZ}
=-2h_{z} c^{(q=0,L)}_{Z(X)^{L-2}Y}.
\label{eq:ZXXXY(L)_relation_k=0}
\end{align}
In a similar manner,
we can show
\begin{align}
-h_{x} c^{(q=0,L)}_{ZY(X)^{L-3}Z}
=-2h_{z} c^{(q=0,L)}_{Z(X)^{L-2}Y}.
\label{eq:ZXXXY(L)_relation_k=L-3}
\end{align}

Furthermore,
if $4\le L\le N/2$,
$c^{(q=0,L)}_{Z(X)^{L-2}Y}$ contributes to
\begin{align}
\begin{array}{rcccc}
 Z&(X)^{L-3-k}&X&(X)^{k}&Y\\
  &           &Z&       & \\ \hline
-Z&(X)^{L-3-k}&Y&(X)^{k}&Y
\end{array},
\end{align}
with $0\le k\le L-4$.
This also has the other contributions
\begin{align}
\begin{array}{rccccc}
 &Y&(X)^{L-4-k}&Y&(X)^{k}&Y\\
Z&Z&           & &       & \\ \hline
Z&X&(X)^{L-4-k}&Y&(X)^{k}&Y
\end{array},
\label{eq:Diagram_YXYXY(L-1)_ZXYXY(L)}
\end{align}
and
\begin{align}
\begin{array}{rcccc}
Z&(X)^{L-3-k}&Y&(X)^{k}&Z\\
 &           & &       &X\\ \hline
Z&(X)^{L-3-k}&Y&(X)^{k}&Y
\end{array}.
\label{eq:Diagram_ZXYXZ(L)_ZXYXY(L)}
\end{align}
Hence, we have
\begin{align}
&&-h_{z} c^{(q=0,L)}_{Z(X)^{L-2}Y}
+J c^{(q=0,L-1)}_{Y(X)^{L-4-k}Y(X)^{k}Y}&
\notag\\
&&+h_{x} c^{(q=0,L)}_{Z(X)^{L-3-k}Y(X)^{k}Z}
&=0.
\label{eq:ZXXXY(L)_relation1_k}
\end{align}
The second term of the LHS of this equation also contributes to
\begin{align}
\begin{array}{rccccc}
Y&(X)^{L-4-k}&Y&(X)^{k}&Y& \\
 &           & &       &Z&Z\\ \hline
Y&(X)^{L-4-k}&Y&(X)^{k}&X&Z
\end{array},
\label{eq:Diagram_YXYXY(L-1)_YXYXXZ(L)}
\end{align}
which has the other contributions:
\begin{align}
\begin{array}{rccccc}
 Y&(X)^{L-4-k}&X&(X)^{k+1}&Z\\
  &           &Z&         & \\ \hline
-Y&(X)^{L-4-k}&Y&(X)^{k+1}&Z
\end{array},
\end{align}
and
\begin{align}
\begin{array}{rccccc}
Z&(X)^{L-4-k}&Y&(X)^{k+1}&Z\\
X&           & &         & \\ \hline
Y&(X)^{L-4-k}&Y&(X)^{k+1}&Z
\end{array}.
\label{eq:Diagram_ZXYXZ(L)_YXYXZ(L)}
\end{align}
Then we have
\begin{align}
&&J c^{(q=0,L-1)}_{Y(X)^{L-4-k}Y(X)^{k}Y}
-h_{z} c^{(q=0,L)}_{Y(X)^{L-2}Z}&
\notag\\
&&+h_{x} c^{(q=0,L)}_{Z(X)^{L-4-k}Y(X)^{k+1}Z}&=0.
\end{align}
Combining this with Eqs.~(\ref{eq:ZX...Y(L)=Y...XZ(L)}) and (\ref{eq:ZXXXY(L)_relation1_k}),
we have
\begin{align}
&h_{x} (c^{(q=0,L)}_{Z(X)^{L-4-k}Y(X)^{k+1}Z}
-c^{(q=0,L)}_{Z(X)^{L-3-k}Y(X)^{k}Z})
\notag\\
&\quad=-2 h_{z} c^{(q=0,L)}_{Z(X)^{L-2}Y}.
\label{eq:ZXXXY(L)_relation_k}
\end{align}
Summing up all Eqs.~(\ref{eq:ZXXXY(L)_relation_k=0}), (\ref{eq:ZXXXY(L)_relation_k=L-3}), and (\ref{eq:ZXXXY(L)_relation_k}) for $k=0,...,L-4$,
we have
\begin{align}
0&=\sum_{k=0}^{L-3}h_{x} (c^{(q=0,L)}_{Z(X)^{L-3-k}Y(X)^{k}Z}
-c^{(q=0,L)}_{Z(X)^{L-3-k}Y(X)^{k}Z})
\\
&=h_{x} c^{(q=0,L)}_{Z(X)^{L-3}YZ}
-h_{x} c^{(q=0,L)}_{ZY(X)^{L-3}Z}
\notag\\
&\quad+\sum_{k=0}^{L-4}h_{x} (c^{(q=0,L)}_{Z(X)^{L-4-k}Y(X)^{k+1}Z}
-c^{(q=0,L)}_{Z(X)^{L-3-k}Y(X)^{k}Z})
\\
&=-2 h_{z}(L-1) c^{(q=0,L)}_{Z(X)^{L-2}Y}.
\end{align}
Thus we obtain the following.\\
\textit{Proposition~B}: 
Assume $J\neq 0$ and $h_{x}\neq 0$.
For $3\le L\le N/2$, 
the solution of Eq.~(\ref{eq:r^q=0}) satisfies
Eq.~(\ref{eq:LemmaB3}) and 
\begin{align}
h_{z} c^{(q=0,L)}_{Z(X)^{L-2}Y}
=h_{z} c^{(q=0,L)}_{Y(X)^{L-2}Z}
=0.
\label{eq:PropB}
\end{align}
(This proposition also includes Prop.~A.)

Hence, if we, in addition, assume $h_{z}\neq 0$,
we have $c^{(q,L)}_{\vb*{A}^{L}}=0$ unless $(A^{1},A^{L})=(Z,Z)$.
Nevertheless,
we do not assume $h_{z}\neq 0$ until Theorem~1 of Sec.~\ref{sec:Proof_bothZ},
so the obtained results can be applied to the transverse-field Ising chain.

\subsection{\label{sec:Proof_bothZ}Coefficients \texorpdfstring{$c^{(q,L)}_{\vb*{A}^{L}}$}{c L} where both ends are $Z$}

Now the remaining coefficients can be written as $c^{(q,L)}_{ZA^2...A^{L-1}Z}$.
We consider their contribution to $r^{(q,L)}_{\vb*{A}^{L}}$.
For $A^2=Z$, 
the following contribution exists:
\begin{align}
\begin{array}{rcccccc}
Z&Z&A^{3}&...&A^{L-1}&Z\\
 & &     &   &       &X\\ \hline
Z&Z&A^{3}&...&A^{L-1}&Y
\end{array}.
\label{eq:Diagram_ZZ...Z(L)_X}
\end{align}
Because both ends of the remaining coefficients $c^{(q,L)}_{\vb*{A}^{L}}$ are $Z$,
Eq.~(\ref{eq:Diagram_ZZ...Z(L)_X}) is the only contribution to $r^{(q,L)}_{ZZA^{3}...A^{L-1}Y}$.
Hence we have
\begin{align}
h_{x} c^{(q,L)}_{ZZA^{3}...A^{L-1}Z}=0.
\end{align}
In a similar manner, we can show\\
\textit{Lemma~C.1}: 
Assume $J\neq 0$ and $h_{x}\neq 0$.
For $3\le L\le N/2$, 
the solution of Eq.~(\ref{eq:r^q=0}) satisfies
\begin{align}
&c^{(q,L)}_{ZA^{2}...A^{L-1}Z}=0\quad \text{for }A^{2}=I,Z\text{ or }A^{L-1}=I,Z.
\label{eq:LemmaC1}
\end{align}
Here the range of $\vb*{A}^{L}=A^{1}...A^{L}$ is given by Eq.~(\ref{eq:Range_A}).

Thus, the remaining coefficients of the form $c^{(q,L)}_{ZA^2...A^{L-1}Z}$
satisfy $A^{2}=X,Y$ and $A^{L-2}=X,Y$.
Next we consider
\begin{align}
\begin{array}{rcccccc}
Z&Y&A^{3}&...&A^{L-1}&Z\\
 & &     &   &       &X\\ \hline
Z&Y&A^{3}&...&A^{L-1}&Y
\end{array}.
\label{eq:Diagram_ZY...Z(L)_X}
\end{align}
Another contribution to $r^{(q,L)}_{ZYA^{3}...A^{L-1}Y}$ is
\begin{align}
\begin{array}{rcccccc}
  &X&A^{3}&...&A^{L-1}&Y\\
 Z&Z&     &   &       & \\ \hline
-Z&Y&A^{3}&...&A^{L-1}&Y
\end{array}.
\label{eq:Diagram_X...Y(L-1)_ZZ}
\end{align}
In addition,
if $A^{3}=...=A^{L-1}=X$
(and $q=0$),
there is the other contribution
\begin{align}
\begin{array}{rcccc}
 Z&X&(X)^{L-3}&Y\\
  &Z&         & \\ \hline
-Z&Y&(X)^{L-3}&Y
\end{array},
\label{eq:Diagram_ZXXXY(L)_ZYXXY(L)}
\end{align}
which, however, vanishes from Eq.~(\ref{eq:PropB}).
As a result, 
we have
\begin{align}
&h_{x} c^{(q,L)}_{ZYA^{3}...A^{L-1}Z}
-J e^{-iq}c^{(q,L-1)}_{XA^{3}...A^{L-1}Y}
=0
\label{eq:ZY...Z(L)_X...Y(L-1)=0}
\end{align}
For the coefficient $c^{(q,L-1)}_{XA^{2}...A^{L-2}Y}$ 
appearing in Eq.~(\ref{eq:ZY...Z(L)_X...Y(L-1)=0}),
we also have
\begin{align}
\begin{array}{rcccccc}
X&A^{2}&...&A^{L-2}&Y& \\
 &     &   &       &Z&Z\\ \hline
X&A^{2}&...&A^{L-2}&X&Z
\end{array}.
\label{eq:Diagram_X...Y(L-1)_ZZ_2}
\end{align}
If $A^{2}=...=A^{L-2}=X$
(and $q=0$),
there is the other contribution to $r^{(q,L)}_{XA^{2}...A^{L-2}XZ}$
\begin{align}
\begin{array}{rcc}
Y&(X)^{L-2}&Z\\
Z&         & \\ \hline
X&(X)^{L-2}&Z
\end{array},
\label{eq:Diagram_YXXXZ(L)_XXXXZ(L)}
\end{align}
which vanishes from Eq.~(\ref{eq:PropB}).
Hence we have
\begin{align}
&J c^{(q,L-1)}_{XA^{2}...A^{L-2}Y}
=0.
\label{eq:X...Y(L-1)=0}
\end{align}

Combining 
Eqs.~(\ref{eq:ZY...Z(L)_X...Y(L-1)=0}) and 
(\ref{eq:X...Y(L-1)=0}), 
we have
\begin{align}
c^{(q,L)}_{ZYA^{3}...A^{L-1}Z}
=0.
\end{align}
In a similar manner,
we can show $c^{(q,L)}_{ZA^{2}...A^{L-2}YZ}=0$.
As a result, we have\\
\textit{Lemma~C.2}: 
Assume $J\neq 0$ and $h_{x}\neq 0$.
For $3\le L\le N/2$, 
the solution of Eq.~(\ref{eq:r^q=0}) satisfies
\begin{align}
&c^{(q,L)}_{ZA^{2}...A^{L-1}Z}=0\quad \text{unless }(A^{2},A^{L-1})=(X,X).
\label{eq:LemmaC2}
\end{align}
Here the range of $\vb*{A}^{L}=A^{1}...A^{L}$ is given by Eq.~(\ref{eq:Range_A}).

Next we examine the remaining coefficients $c^{(q,L)}_{ZXA^{3}...A^{L-2}XZ}$.
From the following diagram,
they contribute to $r^{(q,L)}_{ZXA^{3}...A^{L-2}XY}$,
\begin{align}
\begin{array}{rcccccc}
Z&X&A^{3}&...&A^{L-2}&X&Z\\
 & &     &   &       & &X\\ \hline
Z&X&A^{3}&...&A^{L-2}&X&Y
\end{array},
\label{eq:Diagram_ZX...XZ(L)_ZX...XY(L)}
\end{align}
which also come from
\begin{align}
\begin{array}{rcccccc}
 &Y&A^{3}&...&A^{L-2}&X&Y\\
Z&Z&     &   &       & & \\ \hline
Z&X&A^{3}&...&A^{L-2}&X&Y
\end{array}.
\label{eq:Diagram_Y...XY(L-1)_ZX...XY(L)}
\end{align}
In addition,
if $A^{k}=Y$ for some $k$ ($3\le k\le L-2$)
while other $A^{j}$'s are equal to $X$ (and $q=0$),
there is the contribution to $r^{(q,L)}_{ZXA^{3}...A^{L-2}XY}$
\begin{align}
\begin{array}{rcccccc}
 Z&(X)^{k-2}&X      &(X)^{L-1-k}&Y\\
  &         &Z      &   &       & & \\ \hline
-Z&(X)^{k-2}&A^{k}=Y&(X)^{L-1-k}&Y
\end{array},
\end{align}
which, however, vanishes from Eq.~(\ref{eq:PropB}).
Thus, we have
\begin{align}
h_{x} c^{(q,L)}_{ZXA^{3}...A^{L-2}XZ}
+J e^{-iq}c^{(q,L-1)}_{YA^{3}...A^{L-2}XY}
=0.
\label{eq:ZX...XZ(L)_circulation1}
\end{align}
The second term of the LHS of this equation
contributes to $r^{(q,L)}_{YA^{3}...A^{L-2}XXZ}$
as
\begin{align}
\begin{array}{rcccccc}
Y&A^{3}&...&A^{L-2}&X&Y& \\
 &     &   &       & &Z&Z\\ \hline
Y&A^{3}&...&A^{L-2}&X&X&Z
\end{array}.
\label{eq:Diagram_Y...XY(L-1)_Y...XXZ(L)}
\end{align}
Another contrition to $r^{(q,L)}_{YA^{3}...A^{L-2}XXZ}$ is
\begin{align}
\begin{array}{rcccccc}
Z&A^{3}&...&A^{L-2}&X&X&Z\\
X&     &   &       & & & \\ \hline
Y&A^{3}&...&A^{L-2}&X&X&Z
\end{array}.
\label{eq:Diagram_Z...XXZ(L)_Y...XXZ(L)}
\end{align}
In addition,
if $A^{k}=Y$ for some $k$ ($3\le k\le L-2$)
while other $A^{j}$'s are equal to $X$ (and $q=0$),
there is the other contribution 
\begin{align}
\begin{array}{rcccccc}
 Y&(X)^{k-3}&X      &(X)^{L-k}&Z\\
  &         &Z      &         & \\ \hline
-Y&(X)^{k-3}&A^{k}=Y&(X)^{L-k}&Z
\end{array},
\end{align}
which, however, vanishes from Eq.~(\ref{eq:PropB}).
Thus, we have
\begin{align}
J c^{(q,L-1)}_{YA^{3}...A^{L-2}XY}
+h_{x} c^{(q,L)}_{ZA^{3}...A^{L-2}XXZ}
=0.
\label{eq:ZX...XZ(L)_circulation2}
\end{align}
From Eqs.~(\ref{eq:ZX...XZ(L)_circulation1}) and (\ref{eq:ZX...XZ(L)_circulation2}),
we have
\begin{align}
c^{(q,L)}_{ZXA^{3}...A^{L-2}XZ}
=e^{-iq} c^{(q,L)}_{ZA^{3}...A^{L-2}XXZ}.
\label{eq:ZX...XZ(L)_circulation}
\end{align}
From Lemma~C.$2$,
Eq.~(\ref{eq:ZX...XZ(L)_circulation}) equals to $0$ unless $A^{3}=X$.
By applying Lemma~C.$2$ and Eq.~(\ref{eq:ZX...XZ(L)_circulation}) repeatedly,
we can show\\
\textit{Lemma~C.3}: 
Assume $J\neq 0$ and $h_{x}\neq 0$.
For $3\le L\le N/2$, 
the solution of Eq.~(\ref{eq:r^q=0}) satisfies
\begin{align}
&c^{(q,L)}_{ZA^{2}...A^{L-1}Z}=0\quad 
\text{unless }q=0\text{ and }A^{2}=...=A^{L-1}=X.
\label{eq:LemmaC3}
\end{align}
Here the range of $\vb*{A}^{L}=A^{1}...A^{L}$ is given by Eq.~(\ref{eq:Range_A}).

\subsubsection{Coefficient \texorpdfstring{$c^{(q,L)}_{\vb*{A}^{L}}$}{c L} with \texorpdfstring{$\vb*{A}^{L}=Z(X)^{L-2}Z$}{A=ZXXXZ}}

From Proposition~B and Lemma~C.3,
the remaining coefficients of the form $c^{(q,L)}_{\vb*{A}^{L}}$
are only 
$c^{(q=0,L)}_{Z(X)^{L-2}Y}$, 
$c^{(q=0,L)}_{Y(X)^{L-2}Z}$, and 
$c^{(q=0,L)}_{Z(X)^{L-2}Z}$.
To conclude this subsection,
we examine the coefficients $c^{(q=0,L)}_{Z(X)^{L-2}Z}$.
The following contributions coming from $c^{(q=0,L)}_{Z(X)^{L-2}Z}$ are important:
\begin{align}
\begin{array}{rcc}
Z&(X)^{L-2}&Z\\
X&         & \\ \hline
Y&(X)^{L-2}&Z
\end{array}
\quad
\begin{array}{rccc}
 Z&X&(X)^{L-3}&Z\\
  &Z&         & \\ \hline
-Z&Y&(X)^{L-3}&Z
\end{array}
\notag\\[5pt]
\begin{array}{rcccccc}
 Z&(X)^{k}&X&(X)^{L-3-k}&Z\\
  &       &Z&           & \\ \hline
-Z&(X)^{k}&Y&(X)^{L-3-k}&Z
\end{array}
\quad
\begin{array}{rccc}
 Z&(X)^{L-3}&X&Z\\
  &         &Z& \\ \hline
-Z&(X)^{L-3}&Y&Z
\end{array}.
\label{eq:ImportantDiagrams_ZXXXZ(L)}
\end{align}

For the first diagram,
there is the other contribution,
\begin{align}
\begin{array}{rccc}
Y&(X)^{L-3}&Y& \\
 &         &Z&Z\\ \hline
Y&(X)^{L-3}&X&Z
\end{array},
\end{align}
which results in
\begin{align}
h_{x}c^{(q=0,L)}_{Z(X)^{L-2}Z}
+J c^{(q=0,L-1)}_{Y(X)^{L-3}Y}
=0.
\label{eq:ZXXXZ(L)_1}
\end{align}

For the second diagram of Eq.~(\ref{eq:ImportantDiagrams_ZXXXZ(L)}),
there is another contribution:
\begin{align}
\begin{array}{rccc}
  &X&(X)^{L-3}&Z\\
 Z&Z&         & \\ \hline
-Z&Y&(X)^{L-3}&Z
\end{array}.
\end{align}
Furthermore,
if $L\ge 4$,
there is the other contribution:
\begin{align}
\begin{array}{rccccc}
Z&Y&(X)^{L-4}&Y& \\
 & &         &Z&Z\\ \hline
Z&Y&(X)^{L-4}&X&Z
\end{array}.
\end{align}
Thus, we have
\begin{align}
&&-h_{z}c^{(q=0,L)}_{Z(X)^{L-2}Z}
-J c^{(q=0,L-1)}_{(X)^{L-2}Z}
+J c^{(q=0,L-1)}_{ZY(X)^{L-4}Y}&=0
\notag\\
&&\quad \text{for }4\le L\le N/2.&
\label{eq:ZXXXZ(L)_2}
\end{align}
In a similar manner,
for the fourth diagram of Eq.~(\ref{eq:ImportantDiagrams_ZXXXZ(L)}),
we have
\begin{align}
&&-h_{z}c^{(q=0,L)}_{Z(X)^{L-2}Z}
+J c^{(q=0,L-1)}_{Y(X)^{L-4}YZ}
-J c^{(q=0,L-1)}_{Z(X)^{L-2}}&=0
\notag\\
&&\quad \text{for }4\le L\le N/2.&
\label{eq:ZXXXZ(L)_4}
\end{align}
Note that
for $L=3$,
the second and fourth diagrams are the same,
and we have
\begin{align}
-h_{z}c^{(q=0,3)}_{ZXZ}
-J c^{(q=0,2)}_{XZ}
-J c^{(q=0,2)}_{ZX}
=0.
\label{eq:ZXXXZ(L)_2,4_L=3}
\end{align}

If $L\ge 5$,
the number $k$ of the third diagram
takes $1\le k\le L-4$,
and
there are the other contributions
for the third diagram of Eq.~(\ref{eq:ImportantDiagrams_ZXXXZ(L)}),
\begin{align}
\begin{array}{rcccccc}
 &Y&(X)^{k-1}&Y&(X)^{L-3-k}&Z\\
Z&Z&         & &           & \\ \hline
Z&X&(X)^{k-1}&Y&(X)^{L-3-k}&Z
\end{array}
\\
\begin{array}{rcccccc}
Z&(X)^{k}&Y&(X)^{L-4-k}&Y& \\
 &       & &           &Z&Z\\ \hline
Z&(X)^{k}&Y&(X)^{L-4-k}&X&Z,
\end{array}
\end{align}
which result in
\begin{align}
&&-h_{z}c^{(q=0,L)}_{Z(X)^{L-2}Z}
+J c^{(q=0,L-1)}_{Y(X)^{k-1}Y(X)^{L-3-k}Z}&
\notag\\
&&+J c^{(q=0,L-1)}_{Z(X)^{k}Y(X)^{L-4-k}Y}&=0
\notag\\
&&\text{for }1\le k\le L-4\text{ with }5\le L\le N/2&
\label{eq:ZXXXZ(L)_3}
\end{align}
Note that for $L=4$
the third diagram
coincides with the second or fourth diagram,
and its contribution has already been evaluated
by Eqs.~(\ref{eq:ZXXXZ(L)_2}) and (\ref{eq:ZXXXZ(L)_4}).

Note also that
Eqs.~(\ref{eq:ZXXXZ(L)_1}), (\ref{eq:ZXXXZ(L)_2}), (\ref{eq:ZXXXZ(L)_4}), and (\ref{eq:ZXXXZ(L)_3}) obtained in this section
include
not only $c^{(q=0,L)}_{Z(X)^{L-2}Z}$
but also $c^{(q,L-1)}_{\vb*{A}^{L-1}}$.
To evaluate the former,
we investigate the latter in the next section.

\subsubsection{Coefficients \texorpdfstring{$c^{(q,L-1)}_{\vb*{A}^{L-1}}$}{c L-1}}

In this section,
as mentioned in the end of the previous section,
we examine the coefficients $c^{(q,L-1)}_{\vb*{A}^{L-1}}$
for all $\vb*{A}^{L-1}$
to evaluate $c^{(q=0,L)}_{Z(X)^{L-2}Z}$.

First, we consider 
\begin{align}
\begin{array}{rccccc}
 X&A^{2}&...&A^{L-2}&X& \\
  &     &   &       &Z&Z\\ \hline
-X&A^{2}&...&A^{L-2}&Y&Z
\end{array},
\end{align}
which is the only contribution to
$r^{(q,L)}_{XA^{2}...A^{L-2}YZ}$.
Hence we have
\begin{align}
J c^{(q,L-1)}_{XA^{2}...A^{L-2}X}
=0.
\end{align}
A similar discussion also applies
to 
$c^{(q,L-1)}_{XA^{2}...A^{L-2}Y}$,
$c^{(q,L-1)}_{YA^{2}...A^{L-2}X}$, and
$c^{(q,L-1)}_{YA^{2}...A^{L-2}Y}$, 
because of 
the following fact: 
The contribution to $r^{(q,L)}_{\vb*{A}^{L}}$
with $(A^{1},A^{L})=(X,Z)$, $(Y,Z)$, $(Z,X)$, $(Z,Y)$ 
includes only
$c^{(q,L-1)}_{\vb*{A}^{L-1}}$
and 
a few exceptional $c^{(q,L)}_{\vb*{A}^{L}}$.
The exceptional cases are given by
\begin{gather}
\begin{array}{rccccc}
Z&(X)^{L-2}&Z\\
X&         & \\ \hline
Y&(X)^{L-2}&Z
\end{array}
\quad
\begin{array}{rccccc}
Z&(X)^{L-2}&Z\\
 &         &X\\ \hline
Z&(X)^{L-2}&Y
\end{array}
\label{eq:ExceptionFromZXXXZ(L)}
\\[5pt]
\begin{array}{rccccc}
Z&(X)^{L-2}&Y\\
 &         &Z\\ \hline
Z&(X)^{L-2}&X
\end{array}
\quad
\begin{array}{rccccc}
 Z&(X)^{k}&X&(X)^{L-3-k}&Y\\
  &       &Z&           & \\ \hline
-Z&(X)^{k}&Y&(X)^{L-3-k}&Y
\end{array}
\label{eq:ExceptionFromZXXXY(L)}
\\[5pt]
\begin{array}{rccccc}
Y&(X)^{L-2}&Z\\
Z&         & \\ \hline
X&(X)^{L-2}&Z
\end{array}
\quad
\begin{array}{rccccc}
 Y&(X)^{k}&X&(X)^{L-3-k}&Z\\
  &       &Z&           & \\ \hline
-Y&(X)^{k}&Y&(X)^{L-3-k}&Z
\end{array}.
\label{eq:ExceptionFromYXXXZ(L)}
\end{gather}
Note that 
the contribution of Eqs.~(\ref{eq:ExceptionFromZXXXY(L)}) and (\ref{eq:ExceptionFromYXXXZ(L)})
can be neglected because of Eq.~(\ref{eq:PropB}).
Applying this,
we can show\\
\textit{Lemma~C.4}: 
Assume $J\neq 0$ and $h_{x}\neq 0$.
For $3\le L\le N/2$, 
the solution of Eq.~(\ref{eq:r^q=0}) satisfies
\begin{align}
&c^{(q,L-1)}_{YA^{2}...A^{L-2}Y}=0
\notag\\
&\quad 
\text{unless }q=0\text{ and }A^{2}=...=A^{L-2}=X,
\\
&c^{(q,L-1)}_{XA^{2}...A^{L-2}Y}=0,
\\
&c^{(q,L-1)}_{YA^{2}...A^{L-2}X}=0,
\\
&c^{(q,L-1)}_{XA^{2}...A^{L-2}X}=0.
\label{eq:LemmaC4}
\end{align}
Here the range of $\vb*{A}^{L-1}=A^{1}...A^{L-1}$ is given by Eq.~(\ref{eq:Range_A}).

Then we examine $c^{(q=0,L-1)}_{Y(X)^{L-3}Y}$.
Its contributions to $r^{(q=0,L)}_{\vb*{A}^{L}}$
represented by the following diagrams
are important:
\begin{align}
\begin{array}{rcc}
Y&(X)^{L-3}&Y\\
Z&         & \\ \hline
X&(X)^{L-3}&Y
\end{array}
\quad
\begin{array}{rcc}
Y&(X)^{L-3}&Y\\
 &         &Z\\ \hline
Y&(X)^{L-3}&X
\end{array}
\label{eq:ImportantDiagrams_YXXY(L-1)_1}
\end{align}
and for $0\le k\le L-4$ with $L\ge 4$:
\begin{align}
\begin{array}{rcccccc}
 Y&(X)^{k}&X&(X)^{L-4-k}&Y\\
  &       &Z&           & \\ \hline
-Y&(X)^{k}&Y&(X)^{L-4-k}&Y
\end{array}.
\label{eq:ImportantDiagrams_YXXY(L-1)_2}
\end{align}

For the first diagram of Eq.~(\ref{eq:ImportantDiagrams_YXXY(L-1)_1}),
there is the other contribution:
\begin{align}
\begin{array}{rcc}
X&(X)^{L-3}&Z\\
 &         &X\\ \hline
X&(X)^{L-3}&Y
\end{array}.
\end{align}
(Other contributions are absent because of Lemma~C.4.)
Hence, we have
\begin{align}
h_{z} c^{(q,L-1)}_{Y(X)^{L-3}Y}
+h_{x} c^{(q,L-1)}_{(X)^{L-2}Z}
=0.
\label{eq:YXXY(L-1)_1}
\end{align}

For the second diagram of Eq.~(\ref{eq:ImportantDiagrams_YXXY(L-1)_1}),
we have
\begin{align}
h_{z} c^{(q,L-1)}_{Y(X)^{L-3}Y}
+h_{x} c^{(q,L-1)}_{Z(X)^{L-2}}
=0
\label{eq:YXXY(L-1)_2}
\end{align}
in a similar manner.

For the diagram~(\ref{eq:ImportantDiagrams_YXXY(L-1)_2}),
there are the other contributions:
\begin{align}
\begin{array}{rcccccc}
Y&(X)^{k}&Y&(X)^{L-4-k}&Z\\
 &       & &           &X\\ \hline
Y&(X)^{k}&Y&(X)^{L-4-k}&Y
\end{array}
\notag\\[5pt]
\begin{array}{rcccccc}
Z&(X)^{k}&Y&(X)^{L-4-k}&Y\\
X&       & &           & \\ \hline
Y&(X)^{k}&Y&(X)^{L-4-k}&Y
\end{array}.
\end{align}
Hence, we have
\begin{align}
&&-h_{z} c^{(q,L-1)}_{Y(X)^{L-3}Y}
+h_{x} c^{(q,L-1)}_{Y(X)^{k}Y(X)^{L-4-k}Z}&
\notag\\
&&
+h_{x} c^{(q,L-1)}_{Z(X)^{k}Y(X)^{L-4-k}Y}&=0
\notag\\
&&
\text{for }0\le k\le L-4 \text{ with }4\le L\le N/2.&
\label{eq:YXXY(L-1)_3}
\end{align}

Now we introduce
\begin{align}
S&:=-J c^{(q=0,L-1)}_{(X)^{L-2}Z}-J c^{(q=0,L-1)}_{Z(X)^{L-2}}
\notag\\
&\quad+J\sum_{k=0}^{L-4}
\bigl(
c^{(q=0,L-1)}_{Y(X)^{k}Y(X)^{L-4-k}Z}
+c^{(q=0,L-1)}_{Z(X)^{k}Y(X)^{L-4-k}Y}
\bigr).
\end{align}
(For $L=3$, we define $S=-J c^{(0,2)}_{XZ}-J c^{(0,2)}_{ZX}$.)
From
Eqs.~(\ref{eq:ZXXXZ(L)_2}), (\ref{eq:ZXXXZ(L)_4}), (\ref{eq:ZXXXZ(L)_3}),
we have
\begin{align}
S=&-J c^{(q=0,L-1)}_{(X)^{L-2}Z}
+J c^{(q=0,L-1)}_{ZY(X)^{L-4}Y}
\notag\\
&+J c^{(q=0,L-1)}_{Y(X)^{L-4}YZ}
-J c^{(q=0,L-1)}_{Z(X)^{L-2}}
\notag\\
&+J\sum_{k=1}^{L-4}
\bigl(
c^{(q=0,L-1)}_{Y(X)^{k-1}Y(X)^{L-3-k}Z}
+c^{(q=0,L-1)}_{Z(X)^{k}Y(X)^{L-4-k}Y}
\bigr)
\notag\\
=&(L-2)h_{z} c^{(q=0,L)}_{Z(X)^{L-2}Z}.
\end{align}
On the other hand,
from
Eqs.~(\ref{eq:YXXY(L-1)_1}), (\ref{eq:YXXY(L-1)_2}), (\ref{eq:YXXY(L-1)_3}),
we have
\begin{align}
\frac{h_{x}S}{J}
=&
-h_{x} c^{(q=0,L-1)}_{(X)^{L-2}Z}-h_{x} c^{(q=0,L-1)}_{Z(X)^{L-2}}
\notag\\
&\quad+h_{x}\sum_{k=0}^{L-4}
\bigl(
c^{(q=0,L-1)}_{Y(X)^{k}Y(X)^{L-4-k}Z}
+c^{(q=0,L-1)}_{Z(X)^{k}Y(X)^{L-4-k}Y}
\bigr)
\notag\\
=&(L-1)h_{z} c^{(q=0,L-1)}_{Y(X)^{L-3}Y}.
\end{align}
Combining these,
we have
\begin{align}
h_{z}\bigl(
(L-2)h_{x} c^{(q=0,L)}_{Z(X)^{L-2}Z}
-(L-1)J c^{(q=0,L-1)}_{Y(X)^{L-3}Y}
\bigr)
=0.
\end{align}
Inserting Eq.~(\ref{eq:ZXXXZ(L)_1}) into this,
we obtain
\begin{align}
h_{z}(2L-3)h_{x} c^{(q=0,L)}_{Z(X)^{L-2}Z}
=0.
\end{align}
Thus, we have the following proposition.\\
\textit{Proposition~C}: 
Assume $J\neq 0$ and $h_{x}\neq 0$.
For $3\le L\le N/2$, 
the solution of Eq.~(\ref{eq:r^q=0}) satisfies
Eq.~(\ref{eq:LemmaC3}) and 
\begin{align}
h_{z} c^{(q=0,L)}_{Z(X)^{L-2}Z}
=0.
\label{eq:PropC}
\end{align}

From Propositions~B and C, 
the following theorem immediately follows.\\
\textit{Theorem~1}: 
Assume $J\neq 0$, $h_{x}\neq 0$, and $h_{z}\neq 0$.
For $3\le L\le N/2$, 
the solution of Eq.~(\ref{eq:r^q=0}) satisfies
\begin{align}
c^{(q,L)}_{\vb*{A}^{L}}
=0
\quad\text{for any }\vb*{A}^{L}.
\label{eq:Th1}
\end{align}

\subsection{\label{sec:Proof_L<=2}The case of \texorpdfstring{$L\le 2$}{L<=2}}

In this section,
we analyze the $L$-local conserved quantities with $L=2$.
We show that,
by solving Eq.~(\ref{eq:r^q=0}),
the coefficients $c^{(q,2)}_{\vb*{A}^{2}}$ and $c^{(q,1)}_{\vb*{A}^{1}}$
satisfy Eqs.~(\ref{eq:Result1}) and (\ref{eq:Result2}).
This proof also applies to the case $L=1$
by inserting $c^{(q,2)}_{\vb*{A}^{2}}=0$ into the solution for $L=2$.
Hence, we only consider the case $L=2$.

First, we should remark that Proposition~A and Lemma~B.1, 
i.e., Eqs.~(\ref{eq:PropA}) and (\ref{eq:LemmaB1}),
are also applicable to $L=2$.
Thus, the remaining coefficients are
$c^{(q,2)}_{ZZ}$,
$c^{(q,2)}_{ZY}$,
$c^{(q,2)}_{YZ}$, and
$c^{(q,1)}_{\vb*{A}^{1}}$.

Now we consider the contributions to $r^{(q,2)}_{ZX}$,
\begin{align}
\begin{array}{rcccccc}
Z&Y\\
 &Z\\ \hline
Z&X
\end{array}
\qquad
\begin{array}{rcccccc}
 &Y\\
Z&Z\\ \hline
Z&X
\end{array},
\end{align}
which result in
\begin{align}
h_{z} c^{(q,2)}_{ZY}+J e^{-iq} c^{(q,1)}_{Y}=0.
\end{align}
Similarly, we have
\begin{align}
h_{z} c^{(q,2)}_{YZ}+J c^{(q,1)}_{Y}=0.
\end{align}
On the other hand,
the contribution to $r^{(q,1)}_{Z}$
includes only
\begin{align}
\begin{array}{rcccccc}
 Y\\
 X\\ \hline
-Z
\end{array},
\end{align}
which results in
\begin{align}
h_{x} c^{(q,1)}_{Y}=0.
\end{align}
Thus, we obtain\\
\textit{Lemma~D.1}: 
Assume $J\neq 0$, $h_{x}\neq 0$, and $h_{z}\neq 0$.
For $L=2$,
the solution of Eq.~(\ref{eq:r^q=0}) satisfies
\begin{align}
&c^{(q,\ell)}_{\vb*{A}^{\ell}}=0 \qquad 
\text{for any }c^{(q,\ell)}_{\vb*{A}^{\ell}}
\notag\\
&\text{ other than }
c^{(q=0,0)}_{I}, c^{(q,2)}_{ZZ}, c^{(q,1)}_{Z}, c^{(q,1)}_{X}
\label{eq:LemmaD1}
\end{align}

Finally, we examine 
$c^{(q,2)}_{ZZ}$, $c^{(q,1)}_{Z}$, and $c^{(q,1)}_{X}$.
From the contributions to $r^{(q,1)}_{Y}$,
\begin{align}
\begin{array}{rcccccc}
 X\\
 Z\\ \hline
-Y
\end{array}
\quad
\begin{array}{rcccccc}
Z\\
X\\ \hline
Y
\end{array},
\end{align}
we have
\begin{align}
-h_{z} c^{(q,1)}_{X}
+h_{x} c^{(q,1)}_{Z}
=0.
\end{align}
The contributions to $r^{(q,2)}_{YZ}$,
\begin{align}
\begin{array}{rcccccc}
 X& \\
 Z&Z\\ \hline
-Y&Z
\end{array}
\quad
\begin{array}{rcccccc}
Z&Z\\
X& \\ \hline
Y&Z
\end{array},
\end{align}
give
\begin{align}
-J c^{(q,1)}_{X}
+h_{x} c^{(q,2)}_{ZZ}
=0,
\end{align}
and
the contributions to $r^{(q,2)}_{ZY}$,
\begin{align}
\begin{array}{rcccccc}
  &X\\
 Z&Z\\ \hline
-Z&Y
\end{array}
\quad
\begin{array}{rcccccc}
Z&Z\\
 &X\\ \hline
Z&Y
\end{array},
\end{align}
give
\begin{align}
-J e^{-iq}c^{(q,1)}_{X}
+h_{x} c^{(q,2)}_{ZZ}
=0.
\end{align}
Assuming $J\neq 0$, $h_{x}\neq 0$ and $h_{z}\neq 0$,
we have
\begin{align}
c^{(q,2)}_{ZZ}/J
=c^{(q,1)}_{X}/h_{x}
=c^{(q,1)}_{Z}/h_{z},
\end{align}
and these quantities become zero for $q\neq 0$.
Thus, we obtain the following.\\
\textit{Theorem~2}: 
Assume $J\neq 0$, $h_{x}\neq 0$, and $h_{z}\neq 0$.
For $L=2$, 
the solution of Eq.~(\ref{eq:r^q=0}) satisfies
Eqs.~(\ref{eq:Result1}) and (\ref{eq:Result2}).

We can apply this theorem to the case of $L=1$,
by substituting $c^{(q,2)}_{\vb*{A}^{2}}=0$ into the above solution.
The solution then satisfies
\begin{align}
c^{(q,\ell)}_{\vb*{A}^{\ell}}=0 \qquad 
\text{for any }c^{(q,\ell)}_{\vb*{A}^{\ell}}
\text{ other than }
c^{(q=0,0)}_{I}.
\end{align}

Combining these with Theorem~1 of Sec.~\ref{sec:Proof_bothZ},
we obtain the main result given in Sec.~\ref{sec:Result}.

\section{\label{sec:OBC}Open boundary condition}

In this section, we show that our main result of Sec.~\ref{sec:Result} holds
even for 
the OBC~\footnote{
Other boundary conditions, such as the twisted boundary condition,
can also be treated in a similar manner.
In addition, the analysis of Sec.~\ref{sec:Proof_OBC} can
be extended to the case where the the coupling constants $J,h_{z},h_{x}$ depend on the site $j$~\cite{Imbrie2016}. 
Such a extension will contribute to solving physically important problems such as local conserved quantities of a many-body localized system~\cite{Oganesyan2007,Altman2015,Nandkishore2015,Abanin2019}. 
}.

\subsection{\label{sec:SetupResult_OBC}Setup and result for OBC}

For the OBC, 
the setup explained in Sec.~\ref{sec:Setup}
is changed as follows.
The Hamiltonian for the OBC is given by
\begin{align}
\hat{H}^{\text{OBC}}
=-\sum_{j=1}^{N-1}
J\hat{Z}_{j}\hat{Z}_{j+1}
-\sum_{j=1}^{N}\bigl(
h_{z}\hat{Z}_{j}
+h_{x}\hat{X}_{j}
\bigr).
\label{eq:Hamiltonian^OBC}
\end{align}
Because $\hat{H}^{\text{OBC}}$ acts nontrivially 
only on sites $\{1,2,...,N\}$,
any operators acting outside $\{1,2,...,N\}$
commute with $\hat{H}^{\text{OBC}}$.
To exclude such trivial conserved quantities,
we investigate a local conserved quantity
whose support has an overlap with the sites $\{1,2,...,N\}$,
although we do not assume that its support is included in $\{1,2,...,N\}$.
For this purpose,
we consider operators on the infinite chain $\mathbb{Z}$,
and introduce the $\ell$-support basis operator $\hat{\vb*{A}^{\ell}}_{j}$,
Eqs.~(\ref{eq:basis_A^ell}) and (\ref{eq:Range_A}),
acting on the sites $\{j,j+1,...,j+\ell-1\}\subset\mathbb{Z}$ with $-\ell+2\le j\le N$.
Then our candidate of the $L$-local conserved quantity for the OBC 
is given by
\begin{align}
\hat{Q}^{\text{OBC}}
=\sum_{\ell=1}^{L}
\sum_{j=-\ell+2}^{N}
\sum_{\vb*{A}^{\ell}}
c_{\vb*{A}^{\ell}_{j}}^{(\ell)}
\hat{\vb*{A}^{\ell}}_{j}
+c_{I}
\hat{I}.
\label{eq:L-local_OBC}
\end{align}
Because 
the OBC breaks
the translation invariance,
we analyze the equation $[\hat{Q}^{\text{OBC}},\hat{H}^{\text{OBC}}]=0$ 
without introducing the momentum index $q$ of Sec.~\ref{sec:Setup}.

Our main result given in Sec.~\ref{sec:Result}
can be extended to OBC as follows. (The proof will be given in Sec.~\ref{sec:Proof_OBC}.)
\\[3pt]
\textit{Result for OBC}:
For $J,h_z,h_x\neq 0$
and $L\le N/2$,
there is no $L$-local conserved quantity $\hat{Q}^{\text{OBC}}$
that is linearly independent of $\hat{H}^{\text{OBC}}$ and $\hat{I}$.
In other words,
any solution of $[\hat{Q}^{\text{OBC}},\hat{H}^{\text{OBC}}]=0$
for such $J,h_z,h_x,L$
can be written as
\begin{align}
\hat{Q}^{\text{OBC}}
=c_{I}\hat{I}-c^{(2)}_{Z_{1}Z_{2}}\hat{H}^{\text{OBC}}/J,
\label{eq:Result3_OBC}
\end{align}
where $c_{I}$ and $c^{(2)}_{Z_{1}Z_{2}}$ are arbitrary constants.

This result means that
our main result given in Sec.~\ref{sec:Result}
also holds for the OBC.
In other words, 
the choice of the periodic boundary condition or OBC
does not affect the nonintegrability of the model.
This seems consistent with the fact that
equilibrium statistical mechanics
is insensitive to the choice of the boundary condition.

\subsection{\label{sec:Proof_OBC}Proof for OBC}

In this section,
we prove the ``Result for OBC'' explained in Sec.~\ref{sec:SetupResult_OBC}.

We find that the Result for OBC can be proved 
in almost the same way 
as the proof for the periodic boundary condition given in Sec.~\ref{sec:Proof}.
(More precisely, 
the equations for $c_{\vb*{A}^{\ell}_{j}}^{(\ell)}$ are 
almost the same as those for $c_{\vb*{A}^{\ell}}^{(q=0,\ell)}$ of the periodic boundary case.)
The major difference from the periodic boundary case is
the following two points:
\\
(a)~The candidate $\hat{Q}^{\text{OBC}}$ contains $\hat{\vb*{A}^{\ell}}_{j}$ that acts nontrivially outside the domain $\{1,2,...,N\}$,
that is, $\hat{\vb*{A}^{\ell}}_{j}$ with $j< 1$ and one with $j> N-\ell+1$.
\\
(b)~The contribution from the spin interaction term $J\hat{Z}_{j}\hat{Z}_{j+1}$ 
in the Hamiltonian 
is absent 
when either $j$ or $j+1$ is outside the domain $\{1,2,...,N\}$.
\\
In the following, we mainly focus on these points.

We introduce
coefficients $\{r^{(\ell)}_{\vb*{A}^{\ell}_{j}}\}_{\vb*{A}^{\ell}}$ by
\begin{align}
\frac{1}{2i}[\hat{Q}^{\text{OBC}},-\hat{H}^{\text{OBC}}]
=\sum_{\ell}
\sum_{j}
\sum_{\vb*{A}^{\ell}}
r^{(\ell)}_{\vb*{A}^{\ell}_{j}}
\hat{\vb*{A}^{\ell}}_{j}.
\label{eq:r_OBC}
\end{align}
Our question is to solve
\begin{align}
r^{(\ell)}_{\vb*{A}^{\ell}_{j}}=0
\qquad\text{for all }\vb*{A}^{\ell}_{j}.
\label{eq:r_OBC=0}
\end{align}
To analyze this equation,
we extend the graphical notation Eq.~(\ref{eq:Graphical_ex3}) as follows.
For instance,
if $L\ge 3$, 
the LHS of Eq.~(\ref{eq:r_OBC})
includes
the following term:
\begin{align}
&\frac{1}{2i}[c^{(3)}_{X_{1}Z_{2}Y_{3}}\hat{X}_{1}\hat{Z}_{2}\hat{Y}_{3},
J\sum_{j=1}^{N-1}\hat{Z}_{j}\hat{Z}_{j+1}]\notag\\
&=J c^{(3)}_{X_{1}Z_{2}Y_{3}} 
\frac{1}{2i}
\bigl(
[\hat{X}_{1}\hat{Z}_{2}\hat{Y}_{3},\hat{Z}_{1}\hat{Z}_{2}]
+[\hat{X}_{1}\hat{Z}_{2}\hat{Y}_{3},\hat{Z}_{2}\hat{Z}_{3}]
\notag\\
&\hspace{72pt}
+[\hat{X}_{1}\hat{Z}_{2}\hat{Y}_{3},\hat{Z}_{3}\hat{Z}_{4}]
\bigr)
\label{eq:Graphical_OBC_ex1}
\\
&=J c^{(3)}_{X_{1}Z_{2}Y_{3}} 
\bigl(
-\hat{Y}_{1}\hat{I}_{2}\hat{Y}_{3}
+\hat{X}_{1}\hat{I}_{2}\hat{X}_{3}
+\hat{X}_{1}\hat{Z}_{2}\hat{X}_{3}\hat{Z}_{4}
\bigr).
\label{eq:Graphical_OBC_ex2}
\end{align}
Thus, these three terms contribute
to $r^{(3)}_{Y_{1}I_{2}Y_{3}}$, $r^{(3)}_{X_{1}I_{2}X_{3}}$, and $r^{(4)}_{X_{1}Z_{2}X_{3}Z_{4}}$.
We represent
such contributions by the following diagrams:
\begin{align}
\begin{array}{rcc}
 X_{1}&Z_{2}&Y_{3}\\
 Z_{1}&Z_{2}& \\ \hline
-Y_{1}&I_{2}&Y_{3}
\end{array}
\quad
\begin{array}{rcc}
X_{1}&Z_{2}&Y_{3}\\
     &Z_{2}&Z_{3} \\ \hline
X_{1}&I_{2}&X_{3}
\end{array}
\quad
\begin{array}{rccc}
X_{1}&Z_{2}&Y_{3}& \\
     &     &Z_{3}&Z_{4}\\ \hline
X_{1}&Z_{2}&X_{3}&X_{4}
\end{array}.
\label{eq:Graphical_OBC_ex3}
\end{align}
In each diagram, 
the first row represents the term from $\hat{Q}^{\text{OBC}}$,
the second row the term from $-\hat{H}^{\text{OBC}}$,
and the third row the contribution to $[\hat{Q}^{\text{OBC}},-\hat{H}^{\text{OBC}}]/2i$.

In the following,
we investigate the coefficients $c_{\vb*{A}^{\ell}_{j}}^{(\ell)}$ with $\ell=L$,
and show that 
any solution of Eq.~(\ref{eq:r_OBC=0}) satisfy $c_{\vb*{A}^{L}_{j}}^{(L)}=0$ for $L\ge 3$.
In addition, when $L\le 2$, 
we show that 
any solution of Eq.~(\ref{eq:r_OBC=0})
is of the trivial form of Eq.~(\ref{eq:Result3_OBC}).
The proof is divided into two parts.
The first part examines point~(a),
that is, 
$c_{\vb*{A}^{L}_{j}}^{(L)}$ 
with $j<1$ and $j> N-L+1$.
The second part examines 
$c_{\vb*{A}^{L}_{j}}^{(L)}$ 
with $1\le j\le N-L+1$,
paying attention to point~(b).

\subsubsection{$\hat{\vb*{A}}^{L}_{j}$ that acts nontrivially outside the system}

Now we explain point~(a) mentioned above.
We focus on the coefficients $c_{\vb*{A}^{L}_{j}}^{(L)}$ with ($L\ge 2$ and) $j<1$.
(That is, the left end of $\vb*{A}^{L}_{j}$, $A_{j}^{1}$, is outside the domain $\{1,2,...,N\}$.)
If the right end of $\vb*{A}^{L}$, $A^{L}$, is $Y$, 
then the diagram
\begin{align}
\begin{array}{rcccc}
A_{j}^{1}&A_{j+1}^{2}&...&Y_{j+L-1}& \\
         &           &   &Z_{j+L-1}&Z_{j+L}\\ \hline
A_{j}^{1}&A_{j+1}^{2}&...&X_{j+L-1}&Z_{j+L}
\end{array}
\end{align}
is the only contribution to $r_{A_{j}^{1}A_{j+1}^{2}...X_{j+L-1}Z_{j+L}}^{(L+1)}$.
Assuming $J\neq 0$, we have
\begin{align}
c_{A_{j}^{1}A_{j+1}^{2}...Y_{j+L-1}}^{(L)}=0.
\label{eq:c_InnerEndY}
\end{align}
Moreover,
we can show 
\begin{align}
c_{A_{j}^{1}A_{j+1}^{2}...X_{j+L-1}}^{(L)}=0
\label{eq:c_InnerEndX}
\end{align}
in almost the same manner.
On the other hand,
if the right end of $\vb*{A}^{L}$, $A^{L}$, is $Z$,
we consider the following diagram:
\begin{align}
\begin{array}{rcccc}
A_{j}^{1}&A_{j+1}^{2}&...&Z_{j+L-1}\\
         &           &   &X_{j+L-1}\\ \hline
A_{j}^{1}&A_{j+1}^{2}&...&Y_{j+L-1}
\end{array}.
\end{align}
From Eqs.~(\ref{eq:c_InnerEndY}) and (\ref{eq:c_InnerEndX}),
this diagram is the only contribution to $r_{A_{j}^{1}A_{j+1}^{2}...Y_{j+L-1}}^{(L)}$,
and hence we have
\begin{align}
r_{A_{j}^{1}A_{j+1}^{2}...Y_{j+L-1}}^{(L)}
=h_{x} c_{A_{j}^{1}A_{j+1}^{2}...Z_{j+L-1}}^{(L)}=0.
\label{eq:c_InnerEndZ}
\end{align}
A similar discussion also applies to
the coefficients $c_{\vb*{A}^{L}_{j}}^{(L)}$ with ($L\ge 2$ and) $j>N-L+1$.
Thus, we obtain the following theorem.
\\
\textit{Theorem~1~(OBC)}: 
Assume $J\neq 0$ and $h_{x}\neq 0$.
For $L\ge 2$,
the solution of Eq.~(\ref{eq:r_OBC=0}) satisfies
\begin{align}
c_{\vb*{A}^{L}_{j}}^{(L)}=0\quad
\text{for }j<1, \ j>N-L+1.
\end{align}

Therefore,
the remaining coefficients $c_{\vb*{A}^{L}_{j}}^{(L)}$
are those with $1\le j\le N-L+1$.
In other words, point~(a) mentioned above does not matter
for analysis of the coefficients $c_{\vb*{A}^{L}_{j}}^{(L)}$.

\subsubsection{$\hat{\vb*{A}}^{L}_{j}$ inside the system}

Next we consider point (b) mentioned above.
The equations for the coefficients $c_{\vb*{A}^{L}_{j}}^{(L)}$ with $1< j< N-L+1$
are almost the same as those for the coefficients $c_{\vb*{A}^{L}}^{(q=0,L)}$.
This means that the proof for $c_{\vb*{A}^{L}}^{(q=0,L)}$ given in Sec.~\ref{sec:Proof}
also applies to $c_{\vb*{A}^{L}_{j}}^{(L)}$ with $1< j< N-L+1$.
Therefore,
in the following, 
we mainly focus on
the other coefficients $c_{\vb*{A}^{L}_{j}}^{(L)}$ with $j=1$ and $j=N-L+1$.

We start from the coefficients $c_{\vb*{A}^{L}_{j}}^{(L)}$
where both ends of $\vb*{A}^{L}$ are non-$Z$.
In a similar way as Sec.~\ref{sec:Proof_noZ},
we can obtain the following proposition,
which corresponds to Proposition~A of Sec.~\ref{sec:Proof_noZ}.
\\
\textit{Proposition~A (OBC)}: 
Assume $J\neq 0$. 
For $2\le L\le N/2$,
the solution of Eq.~(\ref{eq:r_OBC=0}) satisfies
\begin{align}
c^{(L)}_{\vb*{A}^{L}_{j}}=0 \quad&\text{ for }A^{1}=X,Y,\text{ and } A^{L}=X,Y,
\label{eq:PropA_OBC}
\end{align}
for $1\le j\le N-L+1$.
Here the range of $\vb*{A}^{L}=A^{1}...A^{L}$ is given by Eq.~(\ref{eq:Range_A}).
\\
In other words, 
any coefficients $c_{\vb*{A}^{L}_{j}}^{(L)}$
whose $A^{1}$ and $A^{L}$ are non-$Z$ are zero.

Next we examine 
the coefficients $c_{\vb*{A}^{L}_{j}}^{(L)}$
where one of $A^{1}$ and $A^{L}$ is $Z$ while the other is non-$Z$.
To show that these coefficients are zero,
we need to modify the proof given in Sec.~\ref{sec:Proof_oneZ}.
For instance,
since the Hamiltonian $\hat{H}^{\text{OBC}}$ does not include the term $J\hat{Z}_{0}\hat{Z}_{1}$,
deriving
\begin{align}
c^{(L)}_{Y_{1}A^{2}_{2}...A^{L-1}_{L-1}Z_{L}}=0\quad \text{for }A^{L-1}=I,Y,Z
\label{eq:Y...IZ(L)=0_OBC}
\end{align}
is not as straightforward as Eq.~(\ref{eq:Y...IZ(L)=0}).
To avoid this difficulty,
we consider the following diagram:
\begin{align}
\begin{array}{rcccc}
Y_{1}&A^{2}_{2}&...&Z_{L}\\
     &         &   &X_{L}\\ \hline
Y_{1}&A^{2}_{2}&...&Y_{L}
\end{array}
\quad
\begin{array}{rcccc}
Z_{1}&A^{2}_{2}&...&Y_{L}\\
X_{1}&         &   & \\ \hline
Y_{1}&A^{2}_{2}&...&Y_{L}
\end{array}.
\end{align}
These are the only contributions to $r^{(L)}_{Y_{1}A^{2}_{2}...A^{L-1}_{L-1}Y_{L}}$,
and hence we have
\begin{align}
c^{(L)}_{Y_{1}A^{2}_{2}...A^{L-1}_{L-1}Z_{L}}
=-c^{(L)}_{Z_{1}A^{2}_{2}...A^{L-1}_{L-1}Y_{L}}.
\label{eq:Y...Z(L)=-Z...Y(L)_OBC}
\end{align}
For the coefficient $c^{(L)}_{Z_{1}A^{2}_{2}...A^{L-1}_{L-1}Y_{L}}$,
we can show that
\begin{align}
c^{(L)}_{Z_{1}A^{2}_{2}...A^{L-1}_{L-1}Y_{L}}=0
\quad\text{unless }A^{2}=...=A^{L-1}=X
\label{eq:ZX...Y(L)=0_OBC}
\end{align}
in a similar manner as Eq.~(\ref{eq:ZX...Y(L)=0}).
Thus we can obtain Eq.~(\ref{eq:Y...IZ(L)=0_OBC}) from 
Eqs.~(\ref{eq:Y...Z(L)=-Z...Y(L)_OBC}) and (\ref{eq:ZX...Y(L)=0_OBC}).
Using these,
we can extend the proof in Sec.~\ref{sec:Proof_oneZ} to OBC,
and obtain the following proposition,
which corresponds to Proposition~B of Sec.~\ref{sec:Proof_oneZ}.
\\
\textit{Proposition~B (OBC)}: 
Assume $J\neq 0$ and $h_{x}\neq 0$.
For $3\le L\le N/2$, 
the solution of Eq.~(\ref{eq:r^q=0}) satisfies
\begin{align}
&c^{(L)}_{\vb*{A}^{L}_{j}}=0\quad &\text{unless }&(A^{1},A^{L})=(Z,Z), 
\notag\\
& & & \vb*{A}^{L}=Z(X)^{L-2}Y,
\notag\\
& & & \vb*{A}^{L}=Y(X)^{L-2}Z,
\end{align}
and 
\begin{align}
&h_{z} c^{(L)}_{\vb*{A}^{L}_{j}}=0\quad &\text{for }&\vb*{A}^{L}=Z(X)^{L-2}Y, 
\notag\\
& & & \vb*{A}^{L}=Y(X)^{L-2}Z,
\label{eq:PropB_OBC}
\end{align}
for $1\le j\le N-L+1$.
Here the range of $\vb*{A}^{L}=A^{1}...A^{L}$ is given by Eq.~(\ref{eq:Range_A}).
\\
In other words, 
if we assume $h_{z}\neq 0$,
then any coefficients $c_{\vb*{A}^{L}_{j}}^{(L)}$
whose $A^{1}$ or $A^{L}$ is non-$Z$ are zero.

Next we examine the coefficients $c^{(L)}_{\vb*{A}^{L}_{j}}$
where both $A^{1}$ and $A^{L}$ are $Z$.
By a discussion similar to above,
we can extend the proof in Sec.~\ref{sec:Proof_bothZ} to the OBC,
and obtain the following proposition,
which corresponds to Proposition~C of Sec.~\ref{sec:Proof_bothZ}.
\\
\textit{Proposition~C (OBC)}: 
Assume $J\neq 0$ and $h_{x}\neq 0$.
For $3\le L\le N/2$, 
the solution of Eq.~(\ref{eq:r_OBC=0}) satisfies
\begin{align}
c^{(L)}_{Z_{j}A^{2}_{j+1}...A^{L-1}_{J+L-2}Z_{j+L-1}}=0\quad 
\text{unless }
A^{2}=...=A^{L-1}=X,
\end{align}
and 
\begin{align}
h_{z} c^{(L)}_{\vb*{A}^{L}_{j}}
=0\quad\text{for }\vb*{A}^{L}=Z(X)^{L-2}Z,
\end{align}
for $1\le j\le N-L+1$.
Here the range of $\vb*{A}^{L}=A^{1}...A^{L}$ is given by Eq.~(\ref{eq:Range_A}).
\\
In other words, 
if we assume $h_{z}\neq 0$,
then any coefficients $c_{\vb*{A}^{L}_{j}}^{(L)}$
whose $A^{1}$ and $A^{L}$ are $Z$ are zero.

Combining Propositions~A~(OBC), B~(OBC), and C~(OBC),
we obtain the following theorem,
which corresponds to Theorem~1 of Sec.~\ref{sec:Proof_bothZ}.
\\
\textit{Theorem~2 (OBC)}: 
Assume $J\neq 0$, $h_{x}\neq 0$, and $h_{z}\neq 0$.
For $3\le L\le N/2$, 
the solution of Eq.~(\ref{eq:r_OBC=0}) satisfies
\begin{align}
c^{(L)}_{\vb*{A}^{L}_{j}}
=0\quad\text{for any }\vb*{A}^{L}
\end{align}
for $1\le j\le N-L+1$.
Here the range of $\vb*{A}^{L}=A^{1}...A^{L}$ is given by Eq.~(\ref{eq:Range_A}).
\\
In other words, 
for $L\ge 3$,
the $L$-local conserved quantity supported on the sites $\{1,2,...,N\}$
does not exist.

In addition,
it is straightforward to extend the proof in Sec.~\ref{sec:Proof_L<=2} to the OBC.
Then we obtain the following theorem,
which corresponds to Theorem~2 of Sec.~\ref{sec:Proof_L<=2}.
\\
\textit{Theorem~3 (OBC)}: 
Assume $J\neq 0$, $h_{x}\neq 0$, and $h_{z}\neq 0$.
For $L\le 2$, 
the solution of Eq.~(\ref{eq:r_OBC=0}) satisfies
\begin{align}
&c^{(\ell)}_{\vb*{A}^{\ell}_{j}}=0 \qquad 
\text{for any }\vb*{A}^{\ell}\text{ and }1\le j\le N-\ell+1
\notag\\
&\text{ other than }
\vb*{A}^{\ell}=ZZ,\  Z,\  X,\  I,
\end{align}
and
\begin{align}
&\frac{c^{(2)}_{Z_{j}Z_{j+1}}}{J}=\frac{c^{(2)}_{Z_{1}Z_{2}}}{J}\quad&
\text{ for }&1\le j\le N-1,
\\
&\frac{c^{(1)}_{Z_{j}}}{h_{z}}=\frac{c^{(1)}_{X_{j}}}{h_{x}}=\frac{c^{(2)}_{Z_{1}Z_{2}}}{J}\quad&
\text{ for }&1\le j\le N.
\end{align}
In other words, 
for $L\le 2$,
any $L$-local conserved quantity supported on the sites $\{1,2,...,N\}$
is restricted to the trivial form of Eq.~(\ref{eq:Result3_OBC}).

From Theorem~1~(OBC), 2~(OBC), 3~(OBC),
we can obtain the Result for OBC, Eq.~(\ref{eq:Result3_OBC}), explained in Sec.~\ref{sec:SetupResult_OBC}.

\section{\label{sec:Discussion} Discussion}

\subsection{\label{sec:Discussion_LargerL} Extension to larger $L$}

In this section,
we discuss why the proof breaks down for $L\ge N/2+1$
and what might be needed to extend our results to such a case.
For simplicity, we assume that $N$ is even in the following of this section.

In the proof of Eq.~(\ref{eq:PropA}),
the existence of the contribution to coefficients $r^{(q,L+1)}_{\vb*{A}^{L+1}}$
from the commutator of $\hat{M}^{(q)}_{\vb*{A}^{L}}$ and $\hat{M}^{(q=0)}_{ZZ}$
is crucial.
However, when $L=N/2+1$, 
there is its exceptional case.
For instance,
there is no contribution from $c^{(q,L)}_{X(I)^{L-2}X}$ to $r^{(q,L+1)}_{\vb*{A}^{L+1}}$
because $\hat{M}^{(q)}_{X(I)^{L-2}YZ}=\hat{M}^{(q)}_{YZ(I)^{L-3}X}$ holds,
as in the diagram
\begin{align}
\begin{array}{rcccccc}
 X&(I)^{L-2}&X& \\
  &         &Z&Z\\ \hline
-X&(I)^{L-2}&Y&Z
\end{array}
\ =\ 
\begin{array}{rcccccc}
 X&I&(I)^{L-3}&X\\
 Z&Z&         & \\ \hline
-Y&Z&(I)^{L-3}&X
\end{array},
\end{align}
due to the periodic boundary condition.
As a result,
it has other contributions, e.g.,
from $c^{(q,L)}_{ZZ(I)^{L-3}X}$:
\begin{align}
\begin{array}{rcccccc}
Z&Z&(I)^{L-3}&X\\
X& &         & \\ \hline
Y&Z&(I)^{L-3}&X
\end{array}.
\end{align}
Thus the proof of Eq.~(\ref{eq:PropA})
and analysis that follows
do not hold.
These terms will require different treatments.

In addition,
for $L\ge N/2+2$,
the additional trivial conserved quantities appear,
and hence 
we need to change
not only the details of the proof,
but also the results themselves.
For example,
the operator $\hat{H}^2$,
which is trivially conserved,
can be represented by
the $L$-local conserved quantity
with $L=N/2+2$.
Hence, our main result
will be extended for $L\ge N/2+2$
as follows:
Any $L$-local conserved quantity $\hat{Q}^{(q)}$ 
can be written as
\begin{align}
\hat{Q}^{(q)}=\delta_{q,0}\bigl(c_{0}\hat{I}+c_{1}\hat{H}+c_{2}\hat{H}^2\bigr).
\end{align}
Furthermore,
if $L$ is larger than $2N/3+2$,
the term proportional to $\hat{H}^{3}$ will be added,
and so forth.
Such extensions will give a key observation 
for investigating 
the presence or absence of quasilocal conserved quantities~\cite{Mierzejewski2015,Ilievski2015,Ilievski2016} of this system.

\subsection{\label{sec:Discussion_TFIM} The transverse-field Ising chain}

In this section, 
we discuss 
the relation between our results
and the model with $h_{z}=0$,
i.e., the transverse-field Ising chain.
First, it is known that
the Hamiltonian of this model
can be mapped to a free fermionic system
by the Jordan-Wigner transformation~\cite{Lieb1961}.
Hence, the conserved quantities of this model
can be easily constructed
in terms of 
the fermionic creation and annihilation operators.
However,
in many cases such conserved quantities are highly nonlocal.
Although it is expected that
suitable linear combinations of
these conserved quantities become local,
it is not so obvious 
what linear combinations would be so.

In such cases,
our results will be useful 
to construct local conserved quantities
of this model directly~\footnote{
It should be mentioned that
the theoretical approach to prove the absence of local conserved quantities 
in a nonintegrable spin chain
developed by Shiraishi~\cite{Shiraishi2019},
on which our proof is based,
has also applied
to integrable models
to explicitly construct local conserved quantities 
by Nozawa and Fukai~\cite{Nozawa2020,Fukai2023,Fukai2023a}.
}.
Since Propositions~B and C of Secs.~\ref{sec:Proof_oneZ} and \ref{sec:Proof_bothZ}
do not assume $h_{z}\neq 0$,
they are also applicable to this model.
These propositions give
the restrictions Eqs.~(\ref{eq:LemmaB3}) and (\ref{eq:LemmaC3})
to the expansion coefficients $c^{(q,\ell)}_{\vb*{A}^{\ell}}$
of $L$-local conserved quantities
with $3\le L\le N/2$.
They say that 
among the coefficients $c^{(q,\ell)}_{\vb*{A}^{\ell}}$
with $\ell=L$,
only
$c^{(q=0,L)}_{Z(X)^{L-2}Z}$ and
$c^{(q=0,L)}_{Z(X)^{L-2}Y}=-c^{(q=0,L)}_{Y(X)^{L-2}Z}$
can be
nonzero.
This statement includes two important implications.
The first is that
there is no $L$-local conserved quantity
for $q\neq 0$ and $3\le L\le N/2$.
The second is that
when constructing $L$-local conserved quantities with $q=0$,
we can set
$c^{(q=0,L)}_{\vb*{A}^{L}}$ to zero
except for the two coefficients mentioned above.

Our second implication mentioned above is consistent with
previous studies on local conserved quantities of the transverse-field Ising chain
as follows.
It is known that this model has Onsager symmetry~\cite{Onsager1944,Minami2021,Miao2022}.
This fact implies that this model has local conserved quantities of the form~\cite{Dolan1982,Klishevich2003}
\begin{align}
\hat{I}^{+}_{L}
&:=
\hat{M}^{(q=0)}_{Z(X)^{L-2}Z}+\hat{M}^{(q=0)}_{Y(X)^{L-4}Y}
\notag\\
&\hspace{12pt}-\frac{h_{x}}{J}(\hat{M}^{(q=0)}_{Z(X)^{L-3}Z}+\hat{M}^{(q=0)}_{Y(X)^{L-3}Y})
\end{align}
for $L>3$, 
and
\begin{align}
\hat{I}^{+}_{L}
:=
\hat{M}^{(q=0)}_{ZXZ}-\hat{M}^{(q=0)}_{X}
-\frac{h_{x}}{J}(\hat{M}^{(q=0)}_{ZZ}+\hat{M}^{(q=0)}_{YY})
\end{align}
for $L=3$.
These expressions are consistent with restrictions Eqs.~(\ref{eq:LemmaB3}) and (\ref{eq:LemmaC3}).
In addition,
it has been shown that this model has other local conserved quantities of the form~\cite{Grady1982,Prosen1998,Fagotti2013}
\begin{align}
\hat{I}^{-}_{L}:=\hat{M}^{(q=0)}_{Z(X)^{L-2}Y}-\hat{M}^{(q=0)}_{Y(X)^{L-2}Z}
\quad \text{for }L\ge 2.
\end{align}
This expression is also consistent with restrictions~(\ref{eq:LemmaB3}) and (\ref{eq:LemmaC3}).
Furthermore, 
our results show that
any $L$-local conserved quantity $\hat{Q}^{(q=0)}$, can be written as
a linear combination of
$\{\hat{I}^{+}_{n},\hat{I}^{-}_{n}\}_{n=3}^{L}$
and some two-local conserved quantity~\footnote{We can also show that any two-local conserved quantity can be written as a linear combination of $\hat{I}^{-}_{2}$, the Hamiltonian, and the identity, by a discussion similar to Sec.~\ref{sec:Proof_L<=2}.}.
This is because, for any $\hat{Q}^{(q=0)}$, 
an appropriate linear combination
$\hat{Q}^{(q=0)}
-c^{(q=0,L)}_{Z(X)^{L-2}Z}\hat{I}^{+}_{L}
-c^{(q=0,L)}_{Z(X)^{L-2}Y}\hat{I}^{-}_{L}$ becomes an $(L-1)$-local conserved quantity.

Moreover,
our main result shows that
the nontrivial conserved quantities $\hat{I}^{+}_{L}$ and $\hat{I}^{-}_{L}$ of the transverse-field Ising chain
are destroyed by any nonzero longitudinal field $h_{z}\neq 0$,
no matter how small $|h_{z}|$ is.
It is interesting that,
even when $|h_{z}|$ is very small,
there is no local conserved quantities
in the neighborhood of $\hat{I}^{+}_{L}$ or $\hat{I}^{-}_{L}$.
The occurrence of such a switching from integrable 
to nonintegrable 
by an arbitrarily small $|h_{z}|$
is considered to be one of the 
mechanisms of prethermalization~\cite{Mori2018,Mallayya2019,Chiba2023}.
Prethermalization 
refers to a nontrivial relaxation dynamics where
the system first relaxes to a nonthermal (quasi)steady state
and then relaxes to the true thermal equilibrium state.

\section{\label{sec:Summary} Summary}

We have proved the absence of local conserved quantities 
in the mixed-field Ising chain Eq.~(\ref{eq:Hamiltonian}) 
with spin interaction $J$, longitudinal field $h_{z}$, and transverse one $h_{x}$.

We have defined an $L$-local conserved quantity
as an operator that is a linear combination of
local operators with support size up to $L$,
and commutes with the Hamiltonian.
Our main result described in Sec.~\ref{sec:Result}
shows that,
if all coupling constants $J,h_{x},h_{z}$ in the Hamiltonian
are nonzero
and $L$ is an integer 
smaller than or equal to 
half of the system size $N/2$,
then any $L$-local conserved quantity
is a trivial one, which
can be written as a linear combination of
the Hamiltonian and the identity.
This result is applicable to the case of the periodic boundary condition.
As mentioned at the end of Sec.~\ref{sec:Result},
this result is fully consistent with 
previously observed chaotic behaviors,
such as the Wigner-Dyson-type level spacing distribution
and the eigenstate thermalization.

This main result has been proved in Sec.~\ref{sec:Proof}
as follows.
Because of the translation invariance of the Hamiltonian $\hat{H}$,
we can restrict a candidate of $L$-local conserved quantities Eq.~(\ref{eq:Q^q})
to an eigenoperator of translation, $\hat{Q}^{(q)}$.
(Here $q$ corresponds to the momentum.)
To express such a quantity efficiently,
we have introduced a basis of eigenoperators of translation, Eq.~(\ref{eq:M^q}).
Each of 
the basis operators
corresponds to a product Eq.~(\ref{eq:basis_A^ell}) of the Pauli operators 
on $\ell$ consecutive sites, $\vb*{A}^{\ell}=A^{1}A^{2}...A^{\ell}$.
Using this basis,
we can expand 
$\hat{Q}^{(q)}$ 
with expansion coefficients 
$c^{(q,\ell)}_{\vb*{A}^{\ell}}$ ($\ell\le L$).
Then we have analyzed the equations for $c^{(q,\ell)}_{\vb*{A}^{\ell}}$
so $\hat{Q}^{(q)}$ commutes with $\hat{H}$.

The analysis of $c^{(q,\ell)}_{\vb*{A}^{\ell}}$ consists of 
four parts (Secs.~\ref{sec:Proof_noZ}--\ref{sec:Proof_L<=2}).
The first to the third parts analyze the coefficients 
with the largest $\ell$, $c^{(q,L)}_{\vb*{A}^{L}}$,
in the case of $3\le L\le N/2$.
The first part (Sec.~\ref{sec:Proof_noZ}) shows that
the coefficients $c^{(q,L)}_{\vb*{A}^{L}}$
whose $A^{1}$ and $A^{L}$ are non-$Z$ are zero.
The second part (Sec.~\ref{sec:Proof_oneZ}) shows that
the coefficients $c^{(q,L)}_{\vb*{A}^{L}}$
where one of $A^{1}$ and $A^{L}$ is $Z$ and the other is non-$Z$ are zero.
The third part (Sec.~\ref{sec:Proof_bothZ}) shows that
the coefficients $c^{(q,L)}_{\vb*{A}^{L}}$
whose $A^{1}$ and $A^{L}$ are both $Z$ are zero.
Thus, these three parts show that there is no $L$-local conserved quantity for $3\le L\le N/2$.
The fourth part (Sec.~\ref{sec:Proof_L<=2}) analyzes the case of $L\le 2$,
and shows that
the coefficients $c^{(q,2)}_{\vb*{A}^{2}}$ and $c^{(q,1)}_{\vb*{A}^{1}}$
are zero except for the trivial ones,
which means that $\hat{Q}^{(q)}$ is a linear combination of $\hat{H}$ and the identity.

Furthermore, we have extended these results
to the case of the OBC in Sec.~\ref{sec:OBC}.
In this case, the translation invariance is absent due to the boundary condition,
and hence, we need to modify the proof.
The key to the extension is the following:
We have taken a site-dependent basis $\hat{\vb*{A}}^{\ell}_{j}$,
which is just the product Eq.~(\ref{eq:basis_A^ell}) of the Pauli matrices,
in place of the basis of eigenoperators of translation, Eq.~(\ref{eq:M^q}).
The coefficients $c^{(\ell)}_{\vb*{A}^{\ell}_{j}}$ in this basis
can be determined from equations almost the same as those for $c^{(q=0,\ell)}_{\vb*{A}^{\ell}}$.

Note that, when analyzing $c^{(q,\ell)}_{\vb*{A}^{\ell}}$,
we have strongly relied on the assumptions $J\neq 0$ and $h_{x}\neq 0$,
but avoided use of $h_{z}\neq 0$ as much as possible.
Therefore, many results in Sec.~\ref{sec:Proof}
are also applicable to the transverse-field Ising chain (the model with $h_{z}=0$).
We have discussed the relation between our results
and the integrability of such a model.

Our results provide a proof of the nonintegrability of 
the mixed-field Ising chain, one of the most well-studied chaotic quantum many-body systems.

\begin{acknowledgments}
We thank N. Shiraishi and A. Shimizu for helpful comments on the paper.
We also thank H. Katsura and R. Hamazaki for suggesting additional references 
on local conserved quantities of the transverse-field Ising chain.
This work is supported by 
Japan Society for the Promotion of Science KAKENHI Grant No. JP21J14313.
\end{acknowledgments}


\bibliography{2024_01_07.bbl}

\end{document}